\documentclass[aps,prl,twocolumn,superscriptaddress,amsmath,amssymb,notitlepage,floatfix]{revtex4-1}
\usepackage[breaklinks=true]{hyperref}
\usepackage[utf8x]{inputenc}
\usepackage{mathtools} 
\usepackage{multirow} 

\hypersetup{
  colorlinks   = true, 
  urlcolor     = blue, 
  linkcolor    = blue, 
  citecolor   =  red 
}

\usepackage{todonotes}

\makeatletter
\def\@bibdataout@aps{%
 \immediate\write\@bibdataout{%
  @CONTROL{%
   apsrev41Control,author="08",editor="1",pages="0",title="0",year="1",eprint="1"%
  }%
 }%
 \if@filesw
  \immediate\write\@auxout{\string\citation{apsrev41Control}}%
 \fi
}%
\makeatother 

\usepackage{graphicx,color}
\usepackage[caption=false]{subfig}
\usepackage{bm} 
\usepackage{dsfont} 
\usepackage{cooltooltips}
\usepackage[capitalise]{cleveref}

\input{shorthand.sty}

\newcommand{\SU}{\mathrm{SU}}
\newcommand{\su}{\mathfrak{su}}
\newcommand{\SO}{\mathrm{SO}}
\newcommand{\bintet}{2\mathrm{T}}
\newcommand{\binoct}{2\mathrm{O}}
\newcommand{\binico}{2\mathrm{I}}

\newcommand{\ketL}[1]{\ket{\bar{#1}}}
\newcommand{\braL}[1]{\bra{\bar{#1}}}
\newcommand{\sigWL}{\overline{\sigma}_w}
\newcommand{\sigXL}{\overline{\sigma}_x}
\newcommand{\sigYL}{\overline{\sigma}_y}
\newcommand{\sigZL}{\overline{\sigma}_z}

\newcommand{\CZ}{\overline{\mathrm{C}{Z}}}
\newcommand{\TL}{\bar{T}}
\newcommand{\SL}{\bar{S}}
\newcommand{\HL}{\bar{H}}
\DeclareMathOperator{\supp}{supp}
\newcommand{\Id}{\mathds{1}}

\begin{document}

\title{Encoding a qubit in a spin}

\author{Jonathan A. Gross}\email{jonathan.gross@usherbrooke.ca}
\affiliation{Institut Quantique \& D\'epartment de Physique, Universit\'e de Sherbrooke, Qu\'ebec J1K 2R1, Canada}

\date{\today}

\begin{abstract}
  I present a new approach for designing quantum error-correcting codes that guarantees a physically natural implementation of Clifford operations.
  Inspired by the scheme put forward by Gottesman, Kitaev, and Preskill for encoding a qubit in an oscillator, in which Clifford operations may be performed via Gaussian unitaries, this approach yields new schemes for encoding a qubit in a large spin in which single-qubit Clifford operations may be performed via spatial rotations.
  I construct all possible examples of such codes, provide universal-gate-set implementations using Hamiltonians that are at most quadratic in angular-momentum operators, and derive criteria for when these codes exactly correct physically relevant noise channels to lowest order, illustrating their performance numerically for specific low-dimensional examples.
\end{abstract}

\maketitle

Great quantum error-correcting codes shield quantum information from a noisy environment while simultaneously making it easily accessible to the programmer.
The very name of these structures betrays an emphasis on the former goal, prioritizing the exact correction of the most likely errors.
In this manuscript I develop an alternative approach to finding new codes that begins by ensuring straightforward logical manipulation of the encoded quantum information.

The encoding of a qubit in an oscillator described by Gottesman, Kitaev, and Preskill~\cite{gottesman_encoding_2001} is an example of a great error-correcting code.
By construction, it protects against unwanted shifts in position and momentum up to a certain threshold.
This protection also optimally corrects damping errors~\cite{noh_quantum_2019}, which are the most prevalent sources of noise in the optical, superconducting, and mechanical systems for which the code is designed.
One can also straightforwardly perform logical operations, since the full set of Clifford operations---the largest set of unitary gates that can be implemented easily---are realized by Hamiltonians at most quadratic in position and momentum---the largest set of Hamiltonians that are easy to engineer in an oscillator.
For these reasons, the Gottesman-Kitaev-Preskill (GKP) code attracts considerable theoretical and experimental attention~\cite{glancy_error_2006,weigand_generating_2018,baragiola_all-gaussian_2019,fluhmann_encoding_2019,campagne-ibarcq_stabilized_2019}.

Other physical systems deserve their own great error-correction codes.
While others have successfully adapted the stabilizer approach of GKP codes to protect against rotational errors~\cite{albert_robust_2019}, alternative single-system codes with easy Cliffords remain unexplored.
I design such codes by starting with an algebra of physical Hamiltonians that are natural to the system at hand.
The construction guarantees that a suitably large and discrete sets of unitary gates---such as logical Clifford operations---can be implemented using only these natural physical interactions.
As a consequence these codes naturally offer resilience against relevant noise channels since environmental fluctuations typically take the form of such natural Hamiltonians.
This approach therefore succeeds in allowing desired manipulations to be performed in a straightforward way while suppressing unwanted environmental interference.

To put this philosophy into practice I demonstrate the construction for large single spins, such as atomic nuclei.
Natural physical operations correspond to spatial rotations of the spin, so I construct all qubit codes on which logical single-qubit Cliffords can be implemented via these spatial rotations.
The codes so constructed allow one to perform the entangling and non Clifford gates necessary for universal quantum computation with only marginally more complex Hamiltonians.
They also automatically exhibit robustness against relevant environmental noise, including random rotations and $T_1$ and $T_2$ processes.
By engineering the satisfaction of a single additional constraint---that the expectation value of $J_z$ vanishes for one of the codewords---these codes exactly correct such environmental noise to lowest order, outperforming all previously studied encodings of qubits into qudits with respect to these errors.
The success of the construction in this particular case builds confidence that the same approach will bear fruit in additional physical systems.

\emph{Encoding qubits in spins.---}The physics of a system dictates which transformations are straightforward.
For large single spins the relevant physics is angular momentum, and the easy transformations are generated by Hamiltonians linear in the angular-momentum operators $J_x$, $J_y$, and $J_z$.
These Hamiltonians arise naturally in practice, for example as the result of driving the spin with a resonant AC magnetic field.
The physical unitaries generated by these Hamiltonians form a representation of the special unitary group $\SU(2)$ on the spin's Hilbert space.
The explicit map from an abstract $\SU(2)$ element to its representative physical unitary is
\begin{align}
  D
  &:
  \exp(-i\theta\hat{\mathbf{n}}\cdot\bm{\sigma}/2)
  \mapsto
  \exp(-i\theta\hat{\mathbf{n}}\cdot\mathbf{J})\,,
\end{align}
where $\bm{\sigma}$ is the vector of abstract Pauli matrices and $\mathbf{J}$ is the vector of the spin's angular-momentum operators.
These representative unitaries are a significantly restricted subgroup of the most general physical unitaries that can act on the large spin's Hilbert space.
Since these restricted unitaries are straightforward to implement, the goal is to find a codespace where the maximum number of logical unitaries can be implemented by physically applying the $\SU(2)$ representatives.

Any $\SU(2)$ representative that realizes a logical unitary must map the codespace to itself.
Because the $\SU(2)$ representation for a large single spin is an irreducible representation (irrep), the only subspaces mapped to themselves by the full set of $\SU(2)$ representatives are the trivial subspace containing only the zero vector and the full Hilbert space of the spin.
Neither of these alternatives is a viable codespace.
The consequence of this observation is that one must limit oneself to a proper subset of $\SU(2)$ representatives when searching for easy physical implementations of logical operations.

I consider two particularly relevant subsets that are representations of finite subgroups of $\SU(2)$.
The subgroup to which I dedicate the most attention is known to quantum-information scientists as the single-qubit Clifford group~\cite{gottesman_heisenberg_1998}, also called the binary octahedral group $\binoct$ because it is the double cover of the rotational symmetry group of the octahedron in the same way $\SU(2)$ is the double cover of $\SO(3)$.
The techniques used for $\binoct$ are easily adapted to other finite subgroups of $\SU(2)$, and I comment on an important example from the binary icosahedral group $\binico$.

For the sake of clarity I now specialize to the subgroup $\binoct$.
The advantage of restricting the set of physical operations to the representatives of $\binoct$ is that these physical operations map nontrivial subspaces to themselves, and these subspaces provide candidate codespaces.
Specifically, the desired qubit codespaces are two-dimensional subspaces of the spin's Hilbert space that are mapped to themselves by $\binoct$ representatives, and on which nontrivial representative unitaries act nontrivially (since the point is for these physical unitaries to act as logical Clifford gates).
In the language of representation theory, the codespaces should be faithful two-dimensional irreps of $\binoct$ obtained by restricting the $\SU(2)$ irrep to the $\binoct$ representatives.

The criteria for the desired codespaces having been established, I now present the representation theory of $\binoct$ needed to establish their existence.

\emph{Identifying binary-octahedral irreps.---}The generators for $\binoct$, concretely realized as $2\times2$ special-unitary matrices, are the phase and Hadamard gates
\begin{align}
  S
  &=
  \exp\big({-}i\tfrac{\pi}{2}\hat{\mathbf{z}}\cdot\bm{\sigma}/2\big)
  =
  \tfrac{1}{\sqrt{2}}(\Id-i\sigma_z)
  \\
  H
  &=
  \exp\big({-}i\pi\tfrac{\hat{\mathbf{x}}+\hat{\mathbf{z}}}{\sqrt{2}}\cdot\bm{\sigma}/2\big)
  =\tfrac{1}{\sqrt{2}}({-}i\sigma_x-i\sigma_z)\,.
\end{align}
The unusual phases are a consequence of the convention to enforce the unit-determinant constraint of special unitaries.
Being a finite group of 48 elements, $\binoct$ possesses only a finite number of irreps.
As detailed in the Supplemental Material~\cite{supp_mat}, only two of these irreps satisfy the criteria of being two dimensional and acting as logical Clifford gates.
Label these two irreps $\varrho_4$ and $\varrho_5$ in recognition of their place amongst the other irreps of $\binoct$.
These irreps are inequivalent as complex representations, $\varrho_4$ straightforwardly mapping $S\mapsto S$ and $H\mapsto H$, but $\varrho_5$ mapping $S\mapsto-S$ and $H\mapsto-H$.
This inequivalence means that codespaces cannot be split between these two irreps, but since the projective action of a unitary $U:\rho\mapsto U\rho U^\dagger$ is all that is relevant from a quantum perspective, the two representations behave identically when considered separately.

\begin{table}
  \centering
  \begin{tabular}{ c c c }
    $\SU(2)$-irrep dim.  & $\varrho_4$ mult. & $\varrho_5$ mult.
    \\
    \hline
    $24q$  & $2q$                                                 & $2q$
    \\
    $24q+2$  & $2q+1$                                               & $2q$
    \\
    $24q+4$  & $2q$                                                 & $2q$
    \\
    $24q+6$  & $2q$                                                 & $2q+1$
    \\
    $24q+8$  & $2q+1$                                               & $2q+1$
    \\
    $24q+10$  & $2q+1$                                               & $2q$
    \\
    $24q+12$  & $2q+1$                                               & $2q+1$
    \\
    $24q+14$  & $2q+1$                                               & $2q+2$
    \\
    $24q+16$  & $2q+1$                                               & $2q+1$
    \\
    $24q+18$  & $2q+2$                                               & $2q+1$
    \\
    $24q+20$ & $2q+2$                                               & $2q+2$
    \\
    $24q+22$ & $2q+1$                                               & $2q+2$
  \end{tabular}
  \caption{Multiplicities of the irreps of interest, $\varrho_4$ and $\varrho_5$, in the reducible $\binoct$ representation derived from the even-dimensional $\SU(2)$ irreps.
  Because these irreps only appear in even dimensions, and their multiplicities follow a pattern that repeats every 24 dimensions, the dimension is presented in the form $24q+2p$, where $q$ is any non-negative integer and $0\leq p\leq11$.}
  \label{tbl:char-inner-prods}
\end{table}

Having identified the two relevant irreps, the task now is to determine whether they appear in the decompositions of the reducible $\binoct$ representations obtained by restricting the $\SU(2)$ irreps to the $\binoct$ representatives.
The decomposition of an irrep of a group into irreps of a subgroup proceeds according to what are called \emph{branching rules}~\cite{fallbacher_breaking_2015}.
The characters of the representations in question are extraordinarily useful in computing such branching rules.
These characters are functions of the group elements obtained by taking traces of the matrices assigned to them by the representation:
\begin{align}
  \chi_D(g)
  &=
  \tr\big(D(g)\big).
\end{align}
Characters of irreps are orthonormal under a suitable inner product, and characters of reducible representations are sums of the characters of the irreps into which they decompose.
Therefore, by taking inner products of reducible characters with the various irrep characters one determines the multiplicity with which each irrep appears in a given reducible representation.

The result of the calculation, worked out explicitly in the Supplemental Material~\cite{supp_mat}, is that the irreps of interest do not appear at all in integer spins (with odd-dimensional Hilbert spaces).
The multiplicities of these irreps in the half-integer spins increase according to a pattern that repeats every 24 dimensions, presented in \cref{tbl:char-inner-prods}.
Spin 1/2 (dimension 2) contains the standard irrep of $\binoct$, but given that this is the entirety of the Hilbert space it does not provide a code.
Spin 3/2 (dimension 4) does not contain any of the irreps of interest, being instead a 4-dimensional irrep of $\binoct$.
For spin 5/2 (dimension 6) and above, however, every half-integer spin contains at least one two-dimensional codespace on which $\binoct$ representatives perform logical Clifford operations.

This result identifies how many codespaces exist in each large single spin.
The next step is to explicitly construct these codes and determine their additional properties.

\begin{figure*}
  \centering
  \subfloat[Spin 5/2, irrep $\varrho_5$]{
    \includegraphics[width=0.45\textwidth]{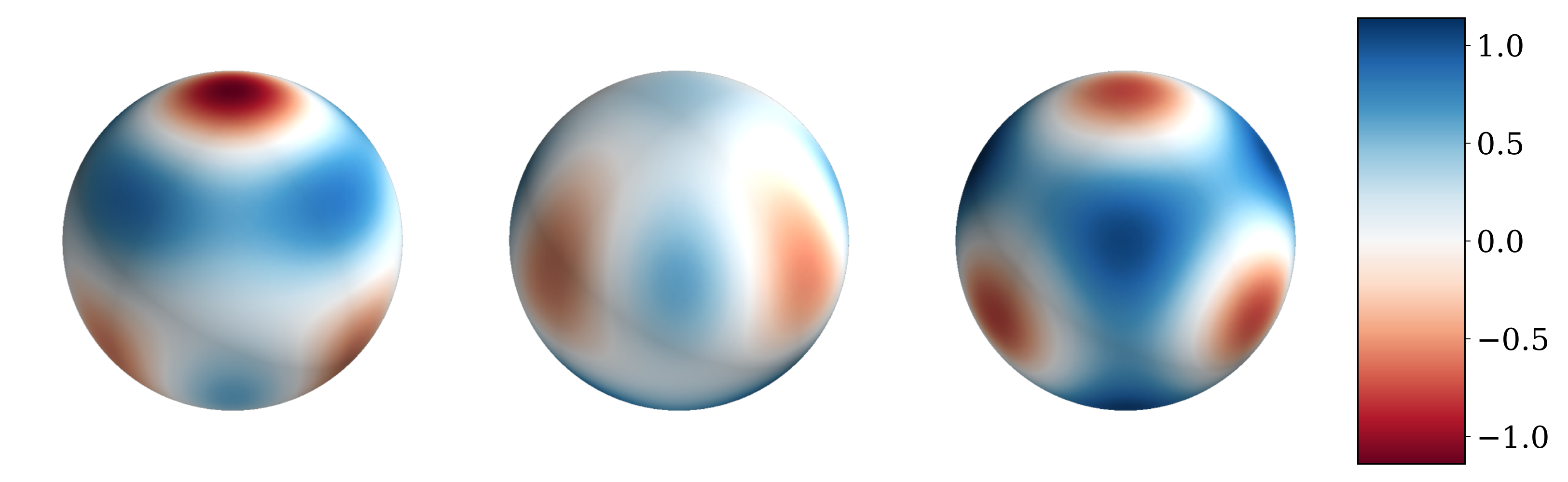}
  }
  ~
  \subfloat[Spin 7/2, irrep $\varrho_5$]{
    \includegraphics[width=0.45\textwidth]{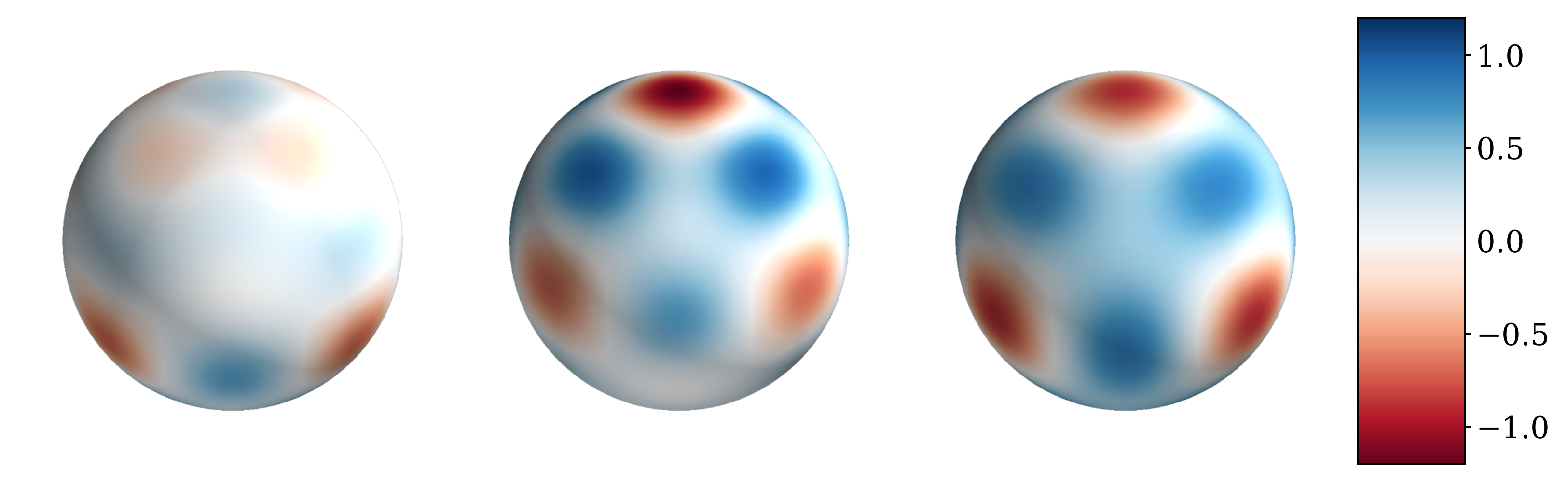}
  }

  \subfloat[Spin 7/2, irrep $\varrho_4$]{
    \includegraphics[width=0.45\textwidth]{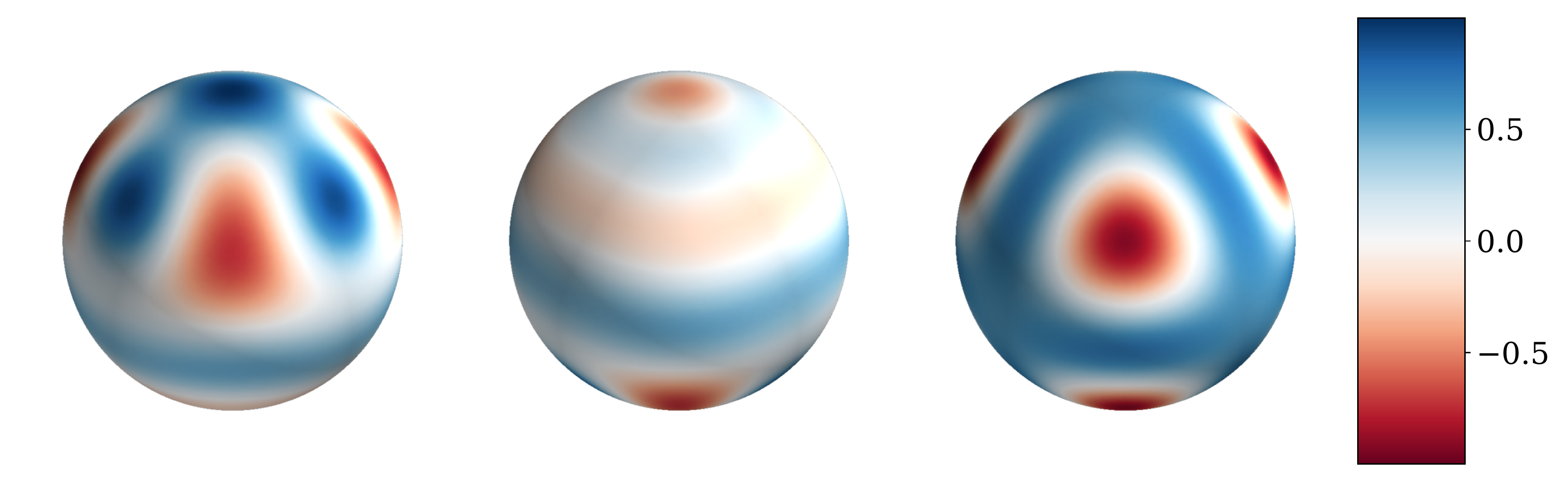}
  }
  ~
  \subfloat[Spin 9/2, irrep $\varrho_4$]{
    \includegraphics[width=0.45\textwidth]{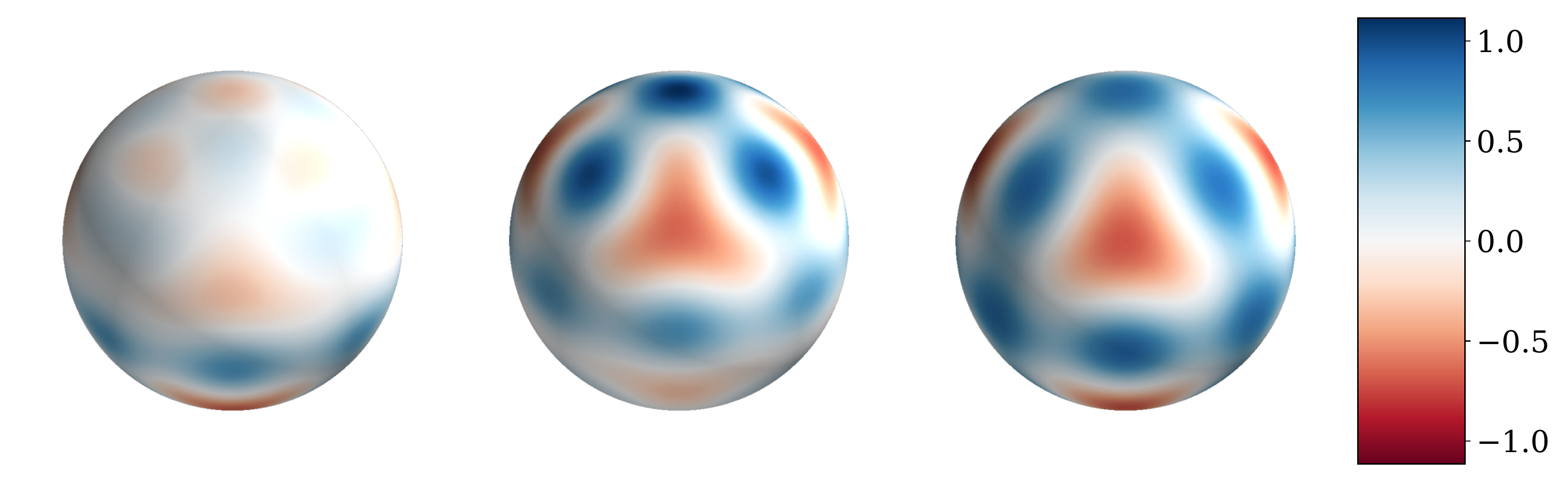}
  }
  \caption{Wigner functions for $\ketL{0}$, $\ketL{1}$, and the codespace projector in order from left to right for several different codes.
  The positive $z$ axis passes through the point of fourfold rotational symmetry at the top of the sphere, and the $x$ and $y$ axis pass through the additional fourfold-rotational-symmetry points of the code projector along the equator.
  Note that, due to the realization of single-qubit Cliffords by $\SU(2)$ rotations, the Wigner function for $\ketL{1}$ is simply the Wigner function for $\ketL{0}$ turned upside down, and the code projector has the symmetry of a cube/octahedron.}
  \label{fig:wigner-fns}
\end{figure*}

\emph{Constructing example codes.---}Producing explicit codewords proceeds by building projectors onto irreps $\varrho_4$ and $\varrho_5$.
The projector onto irrep $\varrho$ of dimension $\dim\varrho$ within reducible representation $D$ emerges as the sum
\begin{align}
  P_{\varrho}
  &=
  \frac{\dim\varrho}{|\binoct|}\sum_{g\in\binoct}\chi_\varrho(g)^*D(g)\,,
\end{align}
where $|\binoct|=48$ is the order of the binary-octahedral group.
The codeword $\ketL{0}$ is taken to be an element of the $+1$ eigenspace of the irrep Pauli $\sigZL$, where irrep Paulis are defined by
\begin{align}\label{eq:codespace-sigz}
  \sigWL\defined P_\varrho\big(i\exp(-i\pi J_w)\big)P_\varrho\,.
\end{align}
To obtain $\ketL{1}$, simply apply $\sigXL$ to $\ketL{0}$.
If the irrep $\varrho$ occurs with multiplicity 1, then the $+1$ eigenspace of $\sigZL$ is one dimensional, and no further choices are required.
If the irrep $\varrho$ occurs with higher multiplicity, further properties of the code can be engineered as explored in the discussion of the error-correction conditions by making an appropriate choice for $\ketL{0}$ within the multidimensional $+1$ eigenspace of $\sigZL$.
\Cref{tbl:low-dimensional-codes} gives the explicit codewords for the four lowest-dimensional codes, each corresponding to an irrep appearing with multiplicity 1.
\Cref{fig:wigner-fns} depicts the Wigner functions for these same codes, defined via a self-dual kernel obeying the Stratonovitch-Weyl postulates for $\SU(2)$~\cite{stratonovich_distributions_1957,heiss_discrete_2000}.
See the Supplemental Material~\cite{supp_mat} for more details.

\begin{table}
  \centering
  \begin{tabular}{ c c r c l }
    Spin & Irrep & \multicolumn{3}{c}{Codewords}
    \\
    \hline
    \\
    \multirow{2}{*}{5/2}
    &
    \multirow{2}{*}{$\varrho_5$}
    &
    $\ketL{0}$
    &
    $=$
    &
    $\sqrt{\tfrac{1}{6}}\,\ketlr{\tfrac{5}{2},\tfrac{5}{2}}
    -\sqrt{\tfrac{5}{6}}\,\ketlr{\tfrac{5}{2},-\tfrac{3}{2}}$
    \\
    & &
    $\ketL{1}$
    &
    $=$
    &
    $-\sqrt{\tfrac{5}{6}}\,\ketlr{\tfrac{5}{2},\tfrac{3}{2}}
    +\sqrt{\tfrac{1}{6}}\,\ketlr{\tfrac{5}{2},-\tfrac{5}{2}}$
    \\
    \\
    \multirow{2}{*}{7/2}
    &
    \multirow{2}{*}{$\varrho_5$}
    &
    $\ketL{0}$
    &
    $=$
    &
    $\tfrac{\sqrt{3}}{2}\,\ketlr{\tfrac{7}{2},\tfrac{5}{2}}
    -\tfrac{1}{2}\,\ketlr{\tfrac{7}{2},-\tfrac{3}{2}}$
    \\
    & &
    $\ketL{1}$
    &
    $=$
    &
    $\tfrac{1}{2}\,\ketlr{\tfrac{7}{2},\tfrac{3}{2}}
    -\tfrac{\sqrt{3}}{2}\,\ketlr{\tfrac{7}{2},-\tfrac{5}{2}}$
    \\
    \\
    \multirow{2}{*}{7/2}
    &
    \multirow{2}{*}{$\varrho_4$}
    &
    $\ketL{0}$
    &
    $=$
    &
    $\sqrt{\tfrac{7}{12}}\,\ketlr{\tfrac{7}{2},\tfrac{1}{2}}
    +\sqrt{\tfrac{5}{12}}\,\ketlr{\tfrac{7}{2},-\tfrac{7}{2}}$
    \\
    & &
    $\ketL{1}$
    &
    $=$
    &
    $-\sqrt{\tfrac{5}{2}}\,\ketlr{\tfrac{7}{2},\tfrac{7}{2}}
    -\sqrt{\tfrac{7}{2}}\,\ketlr{\tfrac{7}{2},-\tfrac{1}{2}}$
    \\
    \\
    \multirow{2}{*}{9/2}
    &
    \multirow{2}{*}{$\varrho_4$}
    &
    $\ketL{0}$
    &
    $=$
    &
    $\tfrac{\sqrt{6}}{4}\,\ketlr{\tfrac{9}{2},\tfrac{9}{2}}
    +\tfrac{\sqrt{21}}{6}\,\ketlr{\tfrac{9}{2},\tfrac{1}{2}}
    +\tfrac{\sqrt{6}}{12}\,\ketlr{\tfrac{9}{2},-\tfrac{7}{2}}$
    \\
    & &
    $\ketL{1}$
    &
    $=$
    &
    $\tfrac{\sqrt{6}}{12}\,\ketlr{\tfrac{9}{2},\tfrac{7}{2}}
    +\tfrac{\sqrt{21}}{6}\,\ketlr{\tfrac{9}{2},-\tfrac{1}{2}}
    +\tfrac{\sqrt{6}}{4}\,\ketlr{\tfrac{9}{2},-\tfrac{9}{2}}$
  \end{tabular}
  \caption{Codewords for the four lowest-dimensional nontrivial examples of $\binoct$-irrep codes.}
  \label{tbl:low-dimensional-codes}
\end{table}

\emph{Computing with encoded qubits.---}Employing these codes in the service of quantum computation requires the ability to do more than single-qubit logical Clifford operations.
I focus now on the following minimal set of logical operations required for universal quantum computation,
\begin{align}
  \{\mathcal{P}_{\ketL{0}},\mathcal{M}_{\sigZL},\SL,\HL,\CZ\}\cup\{\TL\},
\end{align}
where the bars denote logical operators, $\mathcal{P}$ denotes state preparation, and $\mathcal{M}$ denotes operator measurement.
In this set, the single-qubit Cliffords are generated by $\SL$ and $\HL$, multi-qubit Cliffords are obtained by the addition of $\CZ$, and $\TL$ supplies a non Clifford gate.
Since these allow efficient arbitrarily precise approximation of all logical unitaries, the ability to prepare at least one logical state (here chosen to be $\mathcal{P}_{\ketL{0}}$) and perform at least one measurement (here chosen to be $\mathcal{M}_{\sigZL}$) results in universal quantum computation.

By construction these codes have Pauli and single-qubit Clifford operations realizable with Hamiltonians linear in angular-momentum operators (the $\SU(2)$ representation).
This construction gives the codes special structure in the $J_z$ basis which additionally provides explicit recipies for measuring logical Paulis, performing logical $\CZ$ gates between two encoded qubits, and performing logical $\TL$ gates.

To efficiently discuss this structure, I introduce the following notation for a nondegenerate Hermitian operator $A$:
\begin{align}
  \supp_A\ket{\psi}
  &\defined
  \{\lambda\,|\,A\ket{\lambda}=\lambda\ket{\lambda}\,\text{and}\,\iprod{\lambda}{\psi}\neq0\}\,.
\end{align}
This choice leverages the notation for the support of a function, since a vector $\ket{\psi}$ is a function from the eigenbasis of $A$ to the complex numbers and nondegeneracy of $A$ ensures a one-to-one relation between the eigenvectors and eigenvalues.

As shown explicitly in the Supplemental Material~\cite{supp_mat}, restricting to codespaces where $S$ and $X$ are effected by the corresponding $\SU(2)$ representatives implies
\begin{subequations}\label{eq:binoct-supp}
\begin{align}
  \supp_{J_z}\ketL{0}
  &\subseteq
  \begin{cases}
    +\frac{1}{2}+4\mathbf{Z}
  \\
    -\frac{3}{2}+4\mathbf{Z}
  \end{cases}
  \\
  \supp_{J_z}\ketL{1}
  &=
  -\supp_{J_z}\ketL{0}\,,
\end{align}
\end{subequations}
where the term $4\mathbf{Z}$ indicates the set of all integer multiples of 4.
This structure means that one can perform a controlled-$Z$ gate ($\CZ$) using a strategy similar to that used for rotation-symmetric bosonic codes~\cite{grimsmo_quantum_2019}.
In the bosonic case, a cross-Kerr interaction $a^\dagger a\otimes a^\dagger a$ generates the \textsc{crot} gate used to perform $\CZ$ on the codespaces.
In the spin case, the analogous $\tight{J_z\otimes J_z}$ interaction performs the $\CZ$ gate (up to individual $J_z$ corrections).
As worked out in the Supplemental Material~\cite{supp_mat}, the $\CZ$ gate takes the following form:
\begin{align}
  \CZ
  &=
  \exp(i\tfrac{\pi}{2}\tight{J_z\otimes\Id})
  \exp(i\tfrac{\pi}{2}\tight{\Id\otimes J_z})
  \exp(-i\pi\tight{J_z\otimes J_z})\,.
\end{align}

Again, like in rotation-symmetric bosonic codes, a slightly more complicated single-system Hamiltonian yields a more exotic gate.
A self-Kerr interaction $(a^\dagger a)^2$ allows one to perform an $\SL$ gate on the bosonic codes.
The $\binoct$-irrep codes already have an $\SL$ gate using linear Hamiltonians, so adding the analogous $J_z^2$ interaction allows one to perform a $\TL$ gate (again up to a $J_z$ correction).
The $\TL$ gate so obtained, as worked out in the Supplemental Material~\cite{supp_mat}, takes the following forms for the two different $\ketL{0}$ supports:
\begin{align}
  \TL&=\begin{cases}
    \exp(-i\tfrac{\pi}{4}J_z)
    \exp(-i\tfrac{\pi}{4}J_z^2) & m_0=\tfrac{1}{2}
    \\
    \exp(-i\tfrac{5\pi}{4}J_z)
    \exp(-i\tfrac{\pi}{4}J_z^2) & m_0=-\tfrac{3}{2}
  \end{cases}
\end{align}

Destructive measurement in the $\sigZL$ eigenbasis is possible via projecting onto the corresponding angular-momentum basis due to the disjoint support of the eigenstates in these bases.
A nondestructive measurement can be realized with an additional encoded qubit coupled via a $\CZ$ gate which can then be measured destructively as previously described.

Due to the octahedral symmetry of these codes, all the above constructions hold when replacing $z$ with $x$ or $y$.

\emph{Correcting errors.---}As alluded to in the introduction, the fact that only a finite subset of $\SU(2)$ representatives preserve the codespace suggests that these codes might correct errors taking the form of small random $\SU(2)$ representatives in much the same way that GKP codes protect from small random displacements.
I therefore consider noise generated by the Lindblad master equation
\begin{align}
  d\rho
  &=
  \gamma\,dt\sum_{\mathclap{w\in\{x,y,z\}}}(J_w\rho J_w-\tfrac{1}{2}J_w^2\rho-\tfrac{1}{2}\rho J_w^2)\,,
  \label{eq:rand-rot-lind}
\end{align}
where $\gamma$ is the depolarizing rate.
For $\gamma\,dt\ll1$, the following Kraus operators map $\rho\mapsto\rho+d\rho$:
\begin{align}
  E_0
  &=
  \Id-\tfrac{1}{2}\gamma\,dt\,\Vert\mathbf{J}\Vert^2=(1-\tfrac{j(j+1)}{2}\,\gamma\,dt)\Id
  \\
  E_w
  &=
  \sqrt{\gamma\,dt}\,J_w\,.
\end{align}
Correcting the errors corresponding to these Kraus operators is equivalent to correcting random rotations to lowest order.
In spin systems it may be more natural to think of the dominant noise sources in terms of $T_2$-type dephasing errors $J_z$, $T_1$-type relaxation errors $J_-$, and thermalization errors $J_+$.
Since these error operators are linear combinations of the random-rotation error operators, correcting either family of errors is equivalent.
This mirrors the situation in GKP codes, whose manifest protection of random-displacement errors extends to relaxation errors as well~\cite{noh_quantum_2019}.

The elements of the quantum-error-correction matrix indicate whether the codes exactly correct such errors.
The exact-correction condition~\cite{knill_theory_1997} is
\begin{align}
  \braL{a}E_jE_k\ketL{b}
  &=
  C_{jk}\delta_{ab}\,.
\end{align}
Because of the octahedral symmetry of the codes, the conditions reduce to
\begin{subequations}
\begin{align}
  \label{eq:ec-cond-z2}
  \braL{a}J_z^2\ketL{b}
  &=
  C_{zz}\delta_{ab}
  \\
  \label{eq:ec-cond-xy}
  \braL{a}J_xJ_y\ketL{b}
  &=
  C_{xy}\delta_{ab}
  \\
  \label{eq:ec-cond-z}
  \braL{a}J_z\ketL{b}
  &=
  C_{0z}\delta_{ab}\,.
\end{align}
\end{subequations}

Many of these are automatically satisfied by construction.
Because the $\SU(2)$ unitary that inverts $J_z$ also exchanges $\ketL{0}$ and $\ketL{1}$,
\begin{align}
  \braL{1}J_z^2\ketL{1}
  &=
  \braL{0}(-J_z)^2\ketL{0}=\braL{0}J_z^2\ketL{0}\,,
\end{align}
and because $\ketL{0}$ and $\ketL{1}$ have disjoint support on $J_z$,
\begin{align}
  \braL{0}J_z^2\ketL{1}&=\braL{1}J_z^2\ketL{0}=0\,,
\end{align}
completing verification of \cref{eq:ec-cond-z2}.

Since $J_xJ_y\propto J_+^2-2J_z-J_-^2$ and $\braL{a}J_\pm^2\ketL{b}=0$ due to \cref{eq:binoct-supp}, \cref{eq:ec-cond-xy,eq:ec-cond-z} are equivalent to one another.
An additional invokation of the support structure of \cref{eq:binoct-supp} and the $J_z$ inversion yields
\begin{subequations}
\begin{align}
  \braL{0}J_z\ketL{1}
  &=
  \braL{1}J_z\ketL{0}=0
  \\
  \label{eq:jz-antisymmetry}
  \braL{1}J_z\ketL{1}
  &=
  -\braL{0}J_z\ketL{0}\,.
\end{align}
\end{subequations}
The error-correction conditions are therefore satisfied if and only if $\braL{0}J_z\ketL{0}=0$.

In general it is not the case that $\braL{0}J_z\ketL{0}=0$.
For example, in all the codes explicitly presented earlier, $\ketL{0}$ has a nonzero $J_z$ expectation value.
However, if an irrep appears with higher multiplicity, and the projection of $J_z$ onto the $+1$ eigenspace of $\sigZL$ has both positive and negative eigenvalues (or a 0 eigenvalue), then a propitious choice for $\ketL{0}$ ensures that the quantum-error-correction criteria are exactly satisfied for these first-order rotation errors.
The first spin in which one of the irreps appears with higher multiplicity is spin 13/2.
The two eigenvalues of $J_z$ projected onto the $+1$ eigenspace of $\sigZL$ are $-13/6$ and $5/2$, with associated eigenvectors
\begin{align}
  \begin{split}
  \ket{\bar{0}_{-\frac{13}{6}}}
  &=
  \tfrac{\sqrt{910}}{56}\ketlr{\tfrac{13}{2},\tfrac{13}{2}}
  -\tfrac{3\sqrt{154}}{56}\ketlr{\tfrac{13}{2},\tfrac{5}{2}}
  \\
  &\phantrel{=}{}
  -\tfrac{\sqrt{770}}{56}\ketlr{\tfrac{13}{2},-\tfrac{3}{2}}
  +\tfrac{\sqrt{70}}{56}\ketlr{\tfrac{13}{2},-\tfrac{11}{2}}
  \end{split}
  \\
  \begin{split}
  \ket{\bar{0}_{\frac{5}{2}}}
  &=
  \tfrac{\sqrt{231}}{84}\ketlr{\tfrac{13}{2},\tfrac{13}{2}}
  +\tfrac{\sqrt{1365}}{84}\ketlr{\tfrac{13}{2},\tfrac{5}{2}}
  \\
  &\phantrel{=}{}
  -\tfrac{\sqrt{273}}{28}\ketlr{\tfrac{13}{2},-\tfrac{3}{2}}
  -\tfrac{\sqrt{3003}}{84}\ketlr{\tfrac{13}{2},-\tfrac{11}{2}}\,.
  \end{split}
\end{align}
To get a codeword with zero $J_z$ expectation value one takes linear combinations of the following form:
\begin{align}\label{eq:code-13}
  \ket{\bar{0}_\phi}
  &=
  \tfrac{\sqrt{105}}{14}\ket{\bar{0}_{-\frac{13}{6}}}
  +e^{i\phi}\tfrac{\sqrt{91}}{14}\ket{\bar{0}_{\frac{5}{2}}}\,.
\end{align}
Considerations for first-order correction of random-rotation errors make no distinction between different values of the phase $\phi$, leaving a free parameter that may be further optimized over.

Since nuclear spins are obvious host systems for these codes, it would be nice to have examples with good error-correcting properties in a Hilbert space of dimension at most 10 (corresponding to the largest available nuclei of spin 9/2).
The smallest spin with a Hilbert space for large-enough $J_x$, $J_y$, and $J_z$ errors to map the codespace to orthogonal error subspaces is spin 7/2.
As just demonstrated, the $\binoct$ codespaces in spin 7/2 do not have this property.
Using the same tools developed for $\binoct$, one can construct a qubit codespace in spin $7/2$ on which one can use $\SU(2)$ representatives to perform gates corresponding to the symmetries of a regular icosahedron:
\begin{align}
  \ketL{0}
  &=
  \sqrt{\tfrac{3}{10}}\ketlr{\tfrac{7}{2},\tfrac{7}{2}}
  +\sqrt{\tfrac{7}{10}}\ketlr{\tfrac{7}{2},-\tfrac{3}{2}}
  \\
  \ketL{1}
  &=
  \sqrt{\tfrac{7}{10}}\ketlr{\tfrac{7}{2},\tfrac{3}{2}}
  -\sqrt{\tfrac{3}{10}}\ketlr{\tfrac{7}{2},-\tfrac{7}{2}}
\end{align}
Unlike the spin-7/2 $\binoct$ codes, this spin-7/2 binary-icosahedral ($\binico$) code \emph{does} correct small random-rotation errors.
Since the images of the codewords under the various errors span the whole Hilbert space without any overlap, this code is analogous to a perfect block code.

\emph{Measuring code performance.---}To evaluate the performance of these codes under finite-strength random-rotation channels I compute the entanglement fidelity after application of the optimal recovery channel using the semidefinite-programming technique described in~\cite{audenaert_optimizing_2002}.
As a reference I compare to the best previously-considered single-system codes in the given Hilbert space.
These are the minimal qudit codes~\cite{pirandola_minimal_2008} designed to protect against a discrete set of $J_z$-rotation errors and the finite-dimensional analogues of the GKP code~\cite{gottesman_encoding_2001,cafaro_quantum_2012} designed to protect against a discrete set of noncommuting errors.
While the minimal qudit codes are defined for all Hilbert spaces of dimension $4k+2$, The smallest-dimensional example of a qudit GKP code is in spin 17/2.

Note that decoherence-free subspaces and noiseless subsystems for random-rotation errors do not exist in the Hilbert spaces of large single spins since these errors generate an irrep.
Another family of codes designed to protect against rotation errors are molecular codes~\cite{albert_robust_2019}.
In their current formulation, these codes are built in spaces that are direct sums of $\SU(2)$ irreps and additionally protect against shifts in total angular momentum, making direct comparison difficult.

\begin{figure*}
  \centering
  \includegraphics{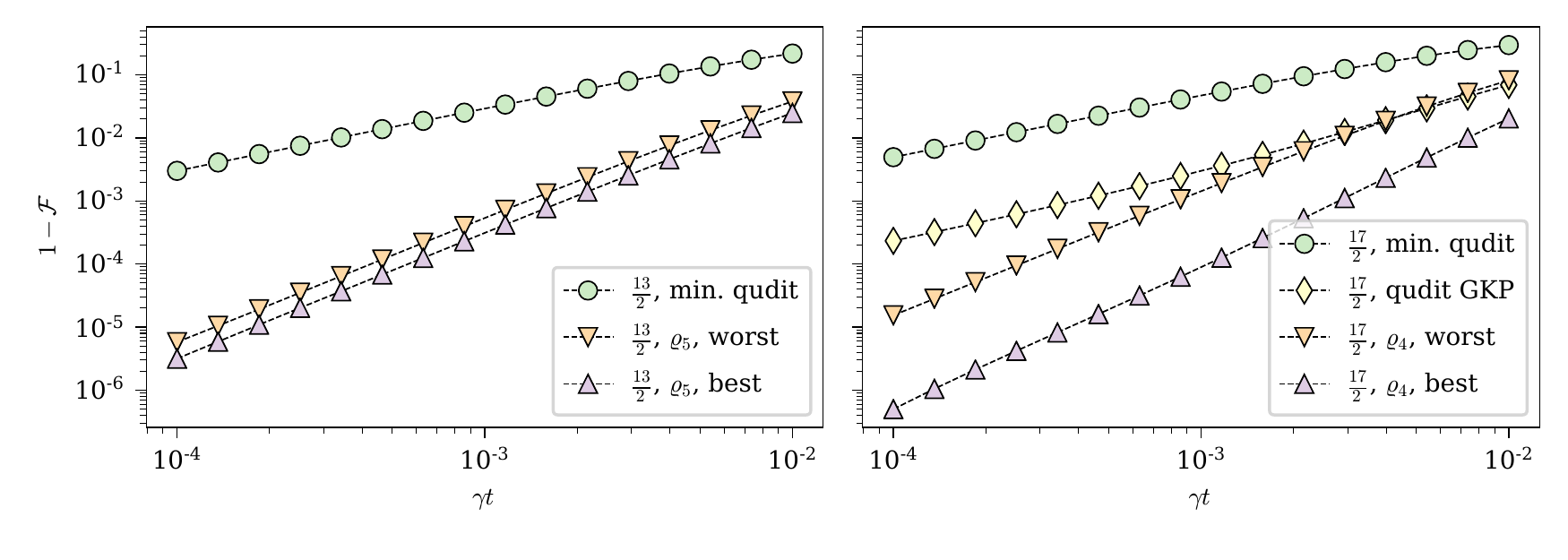}
  \caption{Left: comparison in spin 13/2 of the minimal qudit code (min. qudit) with $\binoct$-irrep codes exactly satisfying the quantum-error-correction conditions for first-order random-rotation errors.
  The first-order correction criteria leave a phase unspecified for the codespace, so both the best and worst choices for the phase are plotted.
  All choices of the undetermined phase exhibit multiple orders of magnitude of improvement over the minimal qudit code, illustrating the power of satisfying the first-order correction criteria.
Right: analogous comparison for spin 17/2 with the addition of the performance of the smallest qudit GKP code (qudit GKP).
  The qudit GKP code is a substantial improvement upon the minimal qudit code, but still dramatically underperforms the $\binoct$-irrep codes which satisfy the quantum-error-correction criteria, regardless of the choice of phase for small-enough evolution times.}
  \label{fig:optimized-fidelity-multiplicity}
\end{figure*}

For small spins where the irreps $\varrho_4$ and $\varrho_5$ appear with multiplicity 1, the nonzero $J_z$ expectation value of the code words results in performance that is only marginally better than that of the minimal qudit codes.
The spin-13/2 codes defined by \cref{eq:code-13} exhibit dramatic improvement over the minimal qudit code, as seen in the left panel of \cref{fig:optimized-fidelity-multiplicity}.

Enlarging the Hilbert space to spin 17/2, the qudit GKP code shows a significant performance increase over the minimal qudit code due to its ability to correct noncommuting errors.
The irrep that appears in spin 17/2 with multiplicity 2 beats all these codes by several orders of magnitude, however, as demonstrated in \cref{fig:optimized-fidelity-multiplicity}.

These simulations demonstrate how well \emph{in principle} this new family of codes can correct against errors, leaving the exact operations required for error correction unspecified.
The highly noncommutitive nature of $\{J_x,J_y,J_z\}$ errors makes the definition of physically natural commuting stabilizers difficult, though one can use the structure of the support in the angular-momentum basis defined in \cref{eq:binoct-supp} to build noncommuting projectors that are analogous to stabilizers.
The construction of practical error-correction procedures using such elements is an ongoing project.

\emph{Generalizing to other systems.---}The construction presented for spin codes exemplifies a more general procedure.
One can replace the representation of the Lie algebra $\su(2)$ given by angular-momentum operators with any representation of a Lie algebra $\mathfrak{g}$ given by physically natural Hamiltonians on a Hilbert space.
Exponentiating these Hamiltonians will generate easily implementable unitaries forming a representation of a Lie group $G$ analogous to $\SU(2)$.
One will then want to consider a discrete subgroup $K\subset G$ just as I considered $\binoct\subset\SU(2)$.
The representation of $G$ restricts to a representation of $K$, and the small-dimensional irreps of $K$ into which this representation decomposes form the candidate codespaces.
At this point one must tailor the procedure to the particular set of errors and the particular discrete subgroup $K$.
When considering random rotations, the error-correction conditions were greatly simplified because the noise was generated by Lindblad operators taken from a subalgebra of $\su(2)$ and $\binoct$ contained a rich set of symmetries of this subalgebra.
One expects similar simplifications to take place in the more general case when analogous structure is present.
Some obvious candidate Lie-algebra representations are those given by quadratic bosonic and fermionic Hamiltonians.
Pursuing the bosonic Hamiltonians brings the prospect of finding additional GKP-like codes in oscillators, though the noncompact nature of the Gaussian unitaries they generate presents qualitatively different challenges than encountered in the $\SU(2)$ case.
Quadratic fermionic Hamiltonians generate compact Lie groups~\cite[Thm.~13.1]{sattinger_lie_2013}, and so provide an arena for a much more straightforward application of the techniques presented here.

\emph{Conclusion.---}In this manuscript I have constructed all single-spin qubit codes admitting Cliffords via $\SU(2)$ unitaries.
These codes exist for all half-integer spins larger than 3/2 and admit the entangling gate $\CZ$ and the non Clifford gate $\TL$ via Hamiltonians quadratic in angular-momentum operators.
I have also exhibited codes in spins as small as 7/2 that exactly protect against random-rotation errors to first order.
In addition to showing how to build better qubits out of large spins, these achievements illustrate the power of the finite-group-representation approach.
Adapting these techniques to systems with different algebras of natural Hamiltonians offers a new path by which to discover useful quantum-error-correcting codes.

I thank Arne L. Grimsmo and Joshua Combes for their insights regarding adapting gate techniques from rotation-symmetric bosonic codes,
Cl\'{e}ment Godfrin and \'{E}va Dupont-Ferrier for inspiring conversations about error correction in spin qubits,
Asaf Diringer and Daniel Carney for helpful discussions about the representation theory used in this construction,
the Les Houches school ``Quantum Information Machines'' for hospitality during a portion of this work,
and Alexandre Blais for guidance and direction throughout the project.
This research was undertaken thanks in part to funding from the Canada First Research Excellence Fund and from NSERC.

\bibliography{references}

\begin{thebibliography}{20}%
\makeatletter
\providecommand \@ifxundefined [1]{%
 \@ifx{#1\undefined}
}%
\providecommand \@ifnum [1]{%
 \ifnum #1\expandafter \@firstoftwo
 \else \expandafter \@secondoftwo
 \fi
}%
\providecommand \@ifx [1]{%
 \ifx #1\expandafter \@firstoftwo
 \else \expandafter \@secondoftwo
 \fi
}%
\providecommand \natexlab [1]{#1}%
\providecommand \enquote  [1]{``#1''}%
\providecommand \bibnamefont  [1]{#1}%
\providecommand \bibfnamefont [1]{#1}%
\providecommand \citenamefont [1]{#1}%
\providecommand \href@noop [0]{\@secondoftwo}%
\providecommand \href [0]{\begingroup \@sanitize@url \@href}%
\providecommand \@href[1]{\@@startlink{#1}\@@href}%
\providecommand \@@href[1]{\endgroup#1\@@endlink}%
\providecommand \@sanitize@url [0]{\catcode `\\12\catcode `\$12\catcode
  `\&12\catcode `\#12\catcode `\^12\catcode `\_12\catcode `\%12\relax}%
\providecommand \@@startlink[1]{}%
\providecommand \@@endlink[0]{}%
\providecommand \url  [0]{\begingroup\@sanitize@url \@url }%
\providecommand \@url [1]{\endgroup\@href {#1}{\urlprefix }}%
\providecommand \urlprefix  [0]{URL }%
\providecommand \Eprint [0]{\href }%
\providecommand \doibase [0]{http://dx.doi.org/}%
\providecommand \selectlanguage [0]{\@gobble}%
\providecommand \bibinfo  [0]{\@secondoftwo}%
\providecommand \bibfield  [0]{\@secondoftwo}%
\providecommand \translation [1]{[#1]}%
\providecommand \BibitemOpen [0]{}%
\providecommand \bibitemStop [0]{}%
\providecommand \bibitemNoStop [0]{.\EOS\space}%
\providecommand \EOS [0]{\spacefactor3000\relax}%
\providecommand \BibitemShut  [1]{\csname bibitem#1\endcsname}%
\let\auto@bib@innerbib\@empty
\bibitem [{\citenamefont {Gottesman}\ \emph {et~al.}(2001)\citenamefont
  {Gottesman}, \citenamefont {Kitaev},\ and\ \citenamefont
  {Preskill}}]{gottesman_encoding_2001}%
  \BibitemOpen
  \bibfield  {author} {\bibinfo {author} {\bibfnamefont {D.}~\bibnamefont
  {Gottesman}}, \bibinfo {author} {\bibfnamefont {A.}~\bibnamefont {Kitaev}}, \
  and\ \bibinfo {author} {\bibfnamefont {J.}~\bibnamefont {Preskill}},\
  }\bibfield  {title} {\enquote {\bibinfo {title} {Encoding a qubit in an
  oscillator},}\ }\href {\doibase 10.1103/PhysRevA.64.012310} {\bibfield
  {journal} {\bibinfo  {journal} {Physical Review A}\ }\textbf {\bibinfo
  {volume} {64}},\ \bibinfo {pages} {012310} (\bibinfo {year}
  {2001})}\BibitemShut {NoStop}%
\bibitem [{\citenamefont {Noh}\ \emph {et~al.}(2019)\citenamefont {Noh},
  \citenamefont {Albert},\ and\ \citenamefont {Jiang}}]{noh_quantum_2019}%
  \BibitemOpen
  \bibfield  {author} {\bibinfo {author} {\bibfnamefont {K.}~\bibnamefont
  {Noh}}, \bibinfo {author} {\bibfnamefont {V.~V.}\ \bibnamefont {Albert}}, \
  and\ \bibinfo {author} {\bibfnamefont {L.}~\bibnamefont {Jiang}},\ }\bibfield
   {title} {\enquote {\bibinfo {title} {Quantum capacity bounds of {Gaussian}
  thermal loss channels and achievable rates with
  {Gottesman}-{Kitaev}-{Preskill} codes},}\ }\href {\doibase
  10.1109/TIT.2018.2873764} {\bibfield  {journal} {\bibinfo  {journal} {IEEE
  Transactions on Information Theory}\ }\textbf {\bibinfo {volume} {65}},\
  \bibinfo {pages} {2563} (\bibinfo {year} {2019})}\BibitemShut {NoStop}%
\bibitem [{\citenamefont {Glancy}\ and\ \citenamefont
  {Knill}(2006)}]{glancy_error_2006}%
  \BibitemOpen
  \bibfield  {author} {\bibinfo {author} {\bibfnamefont {S.}~\bibnamefont
  {Glancy}}\ and\ \bibinfo {author} {\bibfnamefont {E.}~\bibnamefont {Knill}},\
  }\bibfield  {title} {\enquote {\bibinfo {title} {Error analysis for encoding
  a qubit in an oscillator},}\ }\href {\doibase 10.1103/PhysRevA.73.012325}
  {\bibfield  {journal} {\bibinfo  {journal} {Physical Review A}\ }\textbf
  {\bibinfo {volume} {73}},\ \bibinfo {pages} {012325} (\bibinfo {year}
  {2006})}\BibitemShut {NoStop}%
\bibitem [{\citenamefont {Weigand}\ and\ \citenamefont
  {Terhal}(2018)}]{weigand_generating_2018}%
  \BibitemOpen
  \bibfield  {author} {\bibinfo {author} {\bibfnamefont {D.~J.}\ \bibnamefont
  {Weigand}}\ and\ \bibinfo {author} {\bibfnamefont {B.~M.}\ \bibnamefont
  {Terhal}},\ }\bibfield  {title} {\enquote {\bibinfo {title} {Generating grid
  states from {Schr}\"{o}dinger-cat states without postselection},}\ }\href
  {\doibase 10.1103/PhysRevA.97.022341} {\bibfield  {journal} {\bibinfo
  {journal} {Physical Review A}\ }\textbf {\bibinfo {volume} {97}},\ \bibinfo
  {pages} {022341} (\bibinfo {year} {2018})}\BibitemShut {NoStop}%
\bibitem [{\citenamefont {Baragiola}\ \emph {et~al.}(2019)\citenamefont
  {Baragiola}, \citenamefont {Pantaleoni}, \citenamefont {Alexander},
  \citenamefont {Karanjai},\ and\ \citenamefont
  {Menicucci}}]{baragiola_all-gaussian_2019}%
  \BibitemOpen
  \bibfield  {author} {\bibinfo {author} {\bibfnamefont {B.~Q.}\ \bibnamefont
  {Baragiola}}, \bibinfo {author} {\bibfnamefont {G.}~\bibnamefont
  {Pantaleoni}}, \bibinfo {author} {\bibfnamefont {R.~N.}\ \bibnamefont
  {Alexander}}, \bibinfo {author} {\bibfnamefont {A.}~\bibnamefont {Karanjai}},
  \ and\ \bibinfo {author} {\bibfnamefont {N.~C.}\ \bibnamefont {Menicucci}},\
  }\bibfield  {title} {\enquote {\bibinfo {title} {All-{Gaussian} universality
  and fault tolerance with the {Gottesman}-{Kitaev}-{Preskill} code},}\ }\href
  {http://arxiv.org/abs/1903.00012} {\bibfield  {journal} {\bibinfo  {journal}
  {arXiv:1903.00012 [quant-ph]}\ } (\bibinfo {year} {2019})},\ \bibinfo {note}
  {arXiv: 1903.00012}\BibitemShut {NoStop}%
\bibitem [{\citenamefont {Fl\"{u}hmann}\ \emph {et~al.}(2019)\citenamefont
  {Fl\"{u}hmann}, \citenamefont {Nguyen}, \citenamefont {Marinelli},
  \citenamefont {Negnevitsky}, \citenamefont {Mehta},\ and\ \citenamefont
  {Home}}]{fluhmann_encoding_2019}%
  \BibitemOpen
  \bibfield  {author} {\bibinfo {author} {\bibfnamefont {C.}~\bibnamefont
  {Fl\"{u}hmann}}, \bibinfo {author} {\bibfnamefont {T.~L.}\ \bibnamefont
  {Nguyen}}, \bibinfo {author} {\bibfnamefont {M.}~\bibnamefont {Marinelli}},
  \bibinfo {author} {\bibfnamefont {V.}~\bibnamefont {Negnevitsky}}, \bibinfo
  {author} {\bibfnamefont {K.}~\bibnamefont {Mehta}}, \ and\ \bibinfo {author}
  {\bibfnamefont {J.~P.}\ \bibnamefont {Home}},\ }\bibfield  {title} {\enquote
  {\bibinfo {title} {Encoding a qubit in a trapped-ion mechanical
  oscillator},}\ }\href {\doibase 10.1038/s41586-019-0960-6} {\bibfield
  {journal} {\bibinfo  {journal} {Nature}\ }\textbf {\bibinfo {volume} {566}},\
  \bibinfo {pages} {513} (\bibinfo {year} {2019})}\BibitemShut {NoStop}%
\bibitem [{\citenamefont {Campagne-Ibarcq}\ \emph {et~al.}(2019)\citenamefont
  {Campagne-Ibarcq}, \citenamefont {Eickbusch}, \citenamefont {Touzard},
  \citenamefont {Zalys-Geller}, \citenamefont {Frattini}, \citenamefont
  {Sivak}, \citenamefont {Reinhold}, \citenamefont {Puri}, \citenamefont
  {Shankar}, \citenamefont {Schoelkopf}, \citenamefont {Frunzio}, \citenamefont
  {Mirrahimi},\ and\ \citenamefont
  {Devoret}}]{campagne-ibarcq_stabilized_2019}%
  \BibitemOpen
  \bibfield  {author} {\bibinfo {author} {\bibfnamefont {P.}~\bibnamefont
  {Campagne-Ibarcq}}, \bibinfo {author} {\bibfnamefont {A.}~\bibnamefont
  {Eickbusch}}, \bibinfo {author} {\bibfnamefont {S.}~\bibnamefont {Touzard}},
  \bibinfo {author} {\bibfnamefont {E.}~\bibnamefont {Zalys-Geller}}, \bibinfo
  {author} {\bibfnamefont {N.~E.}\ \bibnamefont {Frattini}}, \bibinfo {author}
  {\bibfnamefont {V.~V.}\ \bibnamefont {Sivak}}, \bibinfo {author}
  {\bibfnamefont {P.}~\bibnamefont {Reinhold}}, \bibinfo {author}
  {\bibfnamefont {S.}~\bibnamefont {Puri}}, \bibinfo {author} {\bibfnamefont
  {S.}~\bibnamefont {Shankar}}, \bibinfo {author} {\bibfnamefont {R.~J.}\
  \bibnamefont {Schoelkopf}}, \bibinfo {author} {\bibfnamefont
  {L.}~\bibnamefont {Frunzio}}, \bibinfo {author} {\bibfnamefont
  {M.}~\bibnamefont {Mirrahimi}}, \ and\ \bibinfo {author} {\bibfnamefont
  {M.~H.}\ \bibnamefont {Devoret}},\ }\bibfield  {title} {\enquote {\bibinfo
  {title} {A stabilized logical quantum bit encoded in grid states of a
  superconducting cavity},}\ }\href {http://arxiv.org/abs/1907.12487}
  {\bibfield  {journal} {\bibinfo  {journal} {arXiv:1907.12487 [quant-ph]}\ }
  (\bibinfo {year} {2019})},\ \bibinfo {note} {arXiv: 1907.12487}\BibitemShut
  {NoStop}%
\bibitem [{\citenamefont {Albert}\ \emph {et~al.}(2019)\citenamefont {Albert},
  \citenamefont {Covey},\ and\ \citenamefont {Preskill}}]{albert_robust_2019}%
  \BibitemOpen
  \bibfield  {author} {\bibinfo {author} {\bibfnamefont {V.~V.}\ \bibnamefont
  {Albert}}, \bibinfo {author} {\bibfnamefont {J.~P.}\ \bibnamefont {Covey}}, \
  and\ \bibinfo {author} {\bibfnamefont {J.}~\bibnamefont {Preskill}},\
  }\bibfield  {title} {\enquote {\bibinfo {title} {Robust encoding of a qubit
  in a molecule},}\ }\href {http://arxiv.org/abs/1911.00099} {\bibfield
  {journal} {\bibinfo  {journal} {arXiv:1911.00099 [cond-mat, physics:physics,
  physics:quant-ph]}\ } (\bibinfo {year} {2019})},\ \bibinfo {note} {arXiv:
  1911.00099 version: 1}\BibitemShut {NoStop}%
\bibitem [{\citenamefont {Gottesman}(1998)}]{gottesman_heisenberg_1998}%
  \BibitemOpen
  \bibfield  {author} {\bibinfo {author} {\bibfnamefont {D.}~\bibnamefont
  {Gottesman}},\ }\bibfield  {title} {\enquote {\bibinfo {title} {The
  heisenberg representation of quantum computers},}\ }\href
  {http://arxiv.org/abs/quant-ph/9807006} {\bibfield  {journal} {\bibinfo
  {journal} {arXiv:quant-ph/9807006}\ } (\bibinfo {year} {1998})},\ \bibinfo
  {note} {arXiv: quant-ph/9807006}\BibitemShut {NoStop}%
\bibitem [{sup()}]{supp_mat}%
  \BibitemOpen
  \href@noop {} {}\bibinfo {note} {See supplemental material}\BibitemShut
  {NoStop}%
\bibitem [{\citenamefont {Fallbacher}(2015)}]{fallbacher_breaking_2015}%
  \BibitemOpen
  \bibfield  {author} {\bibinfo {author} {\bibfnamefont {M.}~\bibnamefont
  {Fallbacher}},\ }\bibfield  {title} {\enquote {\bibinfo {title} {Breaking
  classical {Lie} groups to finite subgroups - an automated approach},}\ }\href
  {\doibase 10.1016/j.nuclphysb.2015.07.004} {\bibfield  {journal} {\bibinfo
  {journal} {Nuclear Physics B}\ }\textbf {\bibinfo {volume} {898}},\ \bibinfo
  {pages} {229} (\bibinfo {year} {2015})},\ \bibinfo {note} {arXiv:
  1506.03677}\BibitemShut {NoStop}%
\bibitem [{\citenamefont
  {Stratonovich}(1957)}]{stratonovich_distributions_1957}%
  \BibitemOpen
  \bibfield  {author} {\bibinfo {author} {\bibfnamefont {R.~L.}\ \bibnamefont
  {Stratonovich}},\ }\bibfield  {title} {\enquote {\bibinfo {title} {On
  distributions in representation space},}\ }\href@noop {} {\bibfield
  {journal} {\bibinfo  {journal} {Soviet Physics JETP-USSR}\ }\textbf {\bibinfo
  {volume} {4}},\ \bibinfo {pages} {891} (\bibinfo {year} {1957})}\BibitemShut
  {NoStop}%
\bibitem [{\citenamefont {Heiss}\ and\ \citenamefont
  {Weigert}(2000)}]{heiss_discrete_2000}%
  \BibitemOpen
  \bibfield  {author} {\bibinfo {author} {\bibfnamefont {S.}~\bibnamefont
  {Heiss}}\ and\ \bibinfo {author} {\bibfnamefont {S.}~\bibnamefont
  {Weigert}},\ }\bibfield  {title} {\enquote {\bibinfo {title} {Discrete
  {Moyal}-type representations for a spin},}\ }\href {\doibase
  10.1103/PhysRevA.63.012105} {\bibfield  {journal} {\bibinfo  {journal}
  {Physical Review A}\ }\textbf {\bibinfo {volume} {63}},\ \bibinfo {pages}
  {012105} (\bibinfo {year} {2000})}\BibitemShut {NoStop}%
\bibitem [{\citenamefont {Grimsmo}\ \emph {et~al.}(2019)\citenamefont
  {Grimsmo}, \citenamefont {Combes},\ and\ \citenamefont
  {Baragiola}}]{grimsmo_quantum_2019}%
  \BibitemOpen
  \bibfield  {author} {\bibinfo {author} {\bibfnamefont {A.~L.}\ \bibnamefont
  {Grimsmo}}, \bibinfo {author} {\bibfnamefont {J.}~\bibnamefont {Combes}}, \
  and\ \bibinfo {author} {\bibfnamefont {B.~Q.}\ \bibnamefont {Baragiola}},\
  }\bibfield  {title} {\enquote {\bibinfo {title} {Quantum computing with
  rotation-symmetric bosonic codes},}\ }\href {http://arxiv.org/abs/1901.08071}
  {\bibfield  {journal} {\bibinfo  {journal} {arXiv:1901.08071 [quant-ph]}\ }
  (\bibinfo {year} {2019})},\ \bibinfo {note} {arXiv: 1901.08071}\BibitemShut
  {NoStop}%
\bibitem [{\citenamefont {Knill}\ and\ \citenamefont
  {Laflamme}(1997)}]{knill_theory_1997}%
  \BibitemOpen
  \bibfield  {author} {\bibinfo {author} {\bibfnamefont {E.}~\bibnamefont
  {Knill}}\ and\ \bibinfo {author} {\bibfnamefont {R.}~\bibnamefont
  {Laflamme}},\ }\bibfield  {title} {\enquote {\bibinfo {title} {Theory of
  quantum error-correcting codes},}\ }\href {\doibase 10.1103/PhysRevA.55.900}
  {\bibfield  {journal} {\bibinfo  {journal} {Physical Review A}\ }\textbf
  {\bibinfo {volume} {55}},\ \bibinfo {pages} {900} (\bibinfo {year}
  {1997})}\BibitemShut {NoStop}%
\bibitem [{\citenamefont {Audenaert}\ and\ \citenamefont
  {De~Moor}(2002)}]{audenaert_optimizing_2002}%
  \BibitemOpen
  \bibfield  {author} {\bibinfo {author} {\bibfnamefont {K.}~\bibnamefont
  {Audenaert}}\ and\ \bibinfo {author} {\bibfnamefont {B.}~\bibnamefont
  {De~Moor}},\ }\bibfield  {title} {\enquote {\bibinfo {title} {Optimizing
  completely positive maps using semidefinite programming},}\ }\href {\doibase
  10.1103/PhysRevA.65.030302} {\bibfield  {journal} {\bibinfo  {journal}
  {Physical Review A}\ }\textbf {\bibinfo {volume} {65}},\ \bibinfo {pages}
  {030302} (\bibinfo {year} {2002})}\BibitemShut {NoStop}%
\bibitem [{\citenamefont {Pirandola}\ \emph {et~al.}(2008)\citenamefont
  {Pirandola}, \citenamefont {Mancini}, \citenamefont {Braunstein},\ and\
  \citenamefont {Vitali}}]{pirandola_minimal_2008}%
  \BibitemOpen
  \bibfield  {author} {\bibinfo {author} {\bibfnamefont {S.}~\bibnamefont
  {Pirandola}}, \bibinfo {author} {\bibfnamefont {S.}~\bibnamefont {Mancini}},
  \bibinfo {author} {\bibfnamefont {S.~L.}\ \bibnamefont {Braunstein}}, \ and\
  \bibinfo {author} {\bibfnamefont {D.}~\bibnamefont {Vitali}},\ }\bibfield
  {title} {\enquote {\bibinfo {title} {Minimal qudit code for a qubit in the
  phase-damping channel},}\ }\href {\doibase 10.1103/PhysRevA.77.032309}
  {\bibfield  {journal} {\bibinfo  {journal} {Physical Review A}\ }\textbf
  {\bibinfo {volume} {77}},\ \bibinfo {pages} {032309} (\bibinfo {year}
  {2008})}\BibitemShut {NoStop}%
\bibitem [{\citenamefont {Cafaro}\ \emph {et~al.}(2012)\citenamefont {Cafaro},
  \citenamefont {Maiolini},\ and\ \citenamefont
  {Mancini}}]{cafaro_quantum_2012}%
  \BibitemOpen
  \bibfield  {author} {\bibinfo {author} {\bibfnamefont {C.}~\bibnamefont
  {Cafaro}}, \bibinfo {author} {\bibfnamefont {F.}~\bibnamefont {Maiolini}}, \
  and\ \bibinfo {author} {\bibfnamefont {S.}~\bibnamefont {Mancini}},\
  }\bibfield  {title} {\enquote {\bibinfo {title} {Quantum stabilizer codes
  embedding qubits into qudits},}\ }\href {\doibase 10.1103/PhysRevA.86.022308}
  {\bibfield  {journal} {\bibinfo  {journal} {Physical Review A}\ }\textbf
  {\bibinfo {volume} {86}},\ \bibinfo {pages} {022308} (\bibinfo {year}
  {2012})}\BibitemShut {NoStop}%
\bibitem [{\citenamefont {Sattinger}\ and\ \citenamefont
  {Weaver}(2013)}]{sattinger_lie_2013}%
  \BibitemOpen
  \bibfield  {author} {\bibinfo {author} {\bibfnamefont {D.~H.}\ \bibnamefont
  {Sattinger}}\ and\ \bibinfo {author} {\bibfnamefont {O.~L.}\ \bibnamefont
  {Weaver}},\ }\href@noop {} {\emph {\bibinfo {title} {Lie groups and algebras
  with applications to physics, geometry, and mechanics}}},\ Vol.~\bibinfo
  {volume} {61}\ (\bibinfo  {publisher} {Springer Science \& Business Media},\
  \bibinfo {year} {2013})\BibitemShut {NoStop}%
\bibitem [{\citenamefont {Dokchitser}(2020)}]{dokchitser_groupnames}%
  \BibitemOpen
  \bibfield  {author} {\bibinfo {author} {\bibfnamefont {T.}~\bibnamefont
  {Dokchitser}},\ }\href
  {https://people.maths.bris.ac.uk/~matyd/GroupNames/index.html} {\enquote
  {\bibinfo {title} {{GroupNames}},}\ }\bibinfo {howpublished}
  {\url{https://people.maths.bris.ac.uk/~matyd/GroupNames/index.html}}
  (\bibinfo {year} {2020}),\ \bibinfo {note} {[Online; accessed
  2-April-2020]}\BibitemShut {NoStop}%
\end{thebibliography}%

\newpage
\onecolumngrid
\appendix

\section{\large Supplemental Material}

\section{Irrep multiplicities}

To decompose reducible $\binoct$ representations into their irreps, I make use of the characters of the relevant representations.
The character of a representation $D$ maps group elements to their traces in the representation:
\begin{align}
  \chi_D(g)
  &=
  \tr\big(D(g)\big)\,.
\end{align}
Because the trace doesn't change when you conjugate by an invertible matrix, the character is constant on conjugacy classes $[g]=\{h\,|\,\exists x:g=xhx^{-1}\}$ of the group, and therefore it's efficient to present a character by specifying its values on the conjugacy classes.
The group $\binoct$ has 8 conjugacy classes, which implies that it has only 8 irreps.
Representative elements for each conjugacy class and their images under the characters of the irreps of interest $\varrho_4$ and $\varrho_5$ are presented in \cref{tbl:character}.

Since the reducible representations of interest come from irreps of $\SU(2)$, the Weyl character formula provides the character values, which only depend of the value of $\theta$ in $e^{-i\theta\hat{\mathbf{n}}\cdot\bm{\sigma}/2}$:
\begin{align}
  \chi_{D^{(d)}}(e^{-i\theta\hat{\mathbf{n}}\cdot\bm{\sigma}/2})
  &=
  \frac{\sin d\tfrac{\theta}{2}}{\sin\tfrac{\theta}{2}}\,.
\end{align}
Using the recursive multiple-angle formula $\sin n\theta=2\cos\theta\,\sin(n-1)\theta-\sin(n-2)\theta$ together with $\sin2\theta/\sin\theta=2\cos\theta$ yields a recursive formula for the characters:
\begin{align}
  \chi_{D^{(d)}}
  &=
  \chi_{D^{(2)}}\cdot\chi_{D^{(d-1)}}-\chi_{D^{(d-2)}}\,,
\end{align}
where $\cdot$ denotes pointwise function multiplication: $(f\cdot g)(x)\defined f(x)g(x)$.
The representation $D^{(1)}$ is the trivial representation mapping every group element to the scalar $1$, and corresponds with the $\binoct$ irrep labeled $\varrho_1$.
The representation $D^{(2)}$ is the defining representation for $\SU(2)$, and corresponds with the $\binoct$ irrep labeled $\varrho_4$.
Knowing these characters allows one to compute the characters of all the remaining $\binoct$ representations derived from $\SU(2)$ irreps.
Conjugacy class 1 is the identity representative, which simply yields the value $d$.
Conjugacy class 2 is only slightly more complicated, yielded $(-1)^{d+1}d$.
All other conjugacy classes have character values that are periodic in the dimension, which can be verified as a consequence of the recursive formula by observing the repeated occurrence of two elements in order.
Class 3 repeats with period 3, classes 4a and 4b repeat with period 4, class 6 repeats with period 6, and classes 8a and 8b repeat with period 8, all shown in~\cref{tbl:reducible-character}.
This means all but the first two columns repeat with period 24.

One important property of the characters of the irreps is their orthonormality under the inner product
\begin{align}
  \langle\chi_1,\chi_2\rangle
  &=
  \frac{1}{|G|}\sum_{[g]}|[g]|\chi_1([g])^*\chi_2([g])\,.
\end{align}
Here $|G|$ is the order of the group (number of group elements) and $|[g]|$ is the order of the conjugacy class $[g]$.
Since the character of a reducible representation is the sum of the characters of its irreps, one can count the occurrences of an irrep in a reducible representation by taking the inner product of their characters.

Taking the inner products $\langle\chi_{\varrho_4},\chi_{D^{(d)}}\rangle$ and $\langle\chi_{\varrho_5},\chi_{D^{(d)}}\rangle$ (where $\chi_{D^{(d)}}$ indicates the character of the representation obtained from the $d$-dimensional irrep of $\SU(2)$), I leverage the patterns observed in $\chi_{D^{(d)}}$ to demonstrate that the irreps of interest do not appear in odd dimensions, and obtain the period-24 formulae given in the main text for the $\varrho_4$ and $\varrho_5$ multiplicities in even dimensions.

The same procedure can be carried out for other discrete subgroups of $\SU(2)$.
The double covers of the symmetry groups for the other platonic solids are especially interesting to consider.
These are the binary tetrahedral group $\bintet$ (for the tetrahedron) and the binary icosahedral group $\binico$ (for the icosahedron and its dual, the dodecahedron).
For the sake of completeness I tabulate the characters and multiplicities within the restrictions of $\SU(2)$ irreps of the irreps of $\bintet$, $\binoct$, and $\binico$ in \cref{tbl:2T-2O-char-table,tbl:2I-char-table,tbl:2T-irreps,tbl:2O-irreps,tbl:2I-irreps}.
These tables have been adapted from~\cite{dokchitser_groupnames}, from where the convention for irrep and conjugacy-class labels have also been adopted.

\begin{table*}
  \centering
  \begin{tabular}{ r | c c c c c c c c }
    class & 1 & 2 & 3 & 4a & 4b & 6 & 8a & 8b
    \\
    \hline
    size & 1 & 1 & 8 & 6 & 12 & 8 & 6 & 6
    \\
    rep. elem. & $[\Id]$ & $[-\Id]$
    & $[\tfrac{1}{2}(\tight{-\Id-i\sigma_x-i\sigma_y-i\sigma_z})]$
    & $[-i\sigma_x]$ & $[\tfrac{1}{\sqrt{2}}(\tight{-i\sigma_x-i\sigma_y})]$
    & $[\tfrac{1}{2}(\tight{\Id-i\sigma_x-i\sigma_y-i\sigma_z})]$
    & $[\tfrac{1}{\sqrt{2}}(\tight{\Id-i\sigma_x})]$
    & $[\tfrac{1}{\sqrt{2}}(\tight{-\Id-i\sigma_x})]$
    \\
    $\theta$ & $0$ & $2\pi$ & $\tfrac{4\pi}{3}$ & $\pi$
    & $\pi$ & $\tfrac{2\pi}{3}$ & $\tfrac{\pi}{2}$ & $\tfrac{3\pi}{2}$
    \\
    $\chi_{D^{(1)}}=\chi_{\varrho_1}$ & $1$ & $1$ & $1$ & $1$ & $1$ & $1$ & $1$ & $1$
    \\
    $\chi_{D^{(2)}}=\chi_{\varrho_4}$ & $2$ & $-2$ & $-1$ & $0$ & $0$ & $1$ & $\sqrt{2}$ & $-\sqrt{2}$
    \\
    $\chi_{\varrho_5}$ & $2$ & $-2$ & $-1$ & $0$ & $0$ & $1$ & $-\sqrt{2}$ & $\sqrt{2}$
  \end{tabular}
  \caption{Character-table information for $\binoct$.
  Each column corresponds to a conjugacy class and presents the size (number of elements), a representative element from the class, the rotation angle $\theta$ associated with all elements in the class, and the character value for the three irreps of $\binoct$ needed to calculate the relevant multiplicities: $\varrho_1$, $\varrho_4$, and $\varrho_5$.
  Two of these irreps correspond to $\SU(2)$ irreps restricted to $\binoct$ elements: $D^{(1)}=\varrho_1$ and $D^{(2)}=\varrho_4$.}
  \label{tbl:character}
\end{table*}

\begin{table}
  \centering
  \begin{tabular}{ r | r r r r r r r r }
    class             & 1    & 2     & 3        & 4a       & 4b       & 6        & 8a & 8b
    \\
    \hline
    $\chi_{D^{(1)}}$    & $1$  & $1$   & $1$      & $1$      & $1$      & $1$      & $1$         & $1$
    \\
    $\chi_{D^{(2)}}$    & $2$  & $-2$  & $-1$     & $0$      & $0$      & $0$      & $\sqrt{2}$  & $-\sqrt{2}$
    \\
    $\chi_{D^{(3)}}$    & $3$  & $3$   & $0$      & $-1$     & $-1$     & $-1$     & $1$         & $1$
    \\
    $\chi_{D^{(4)}}$    & $4$  & $-4$  & $1$      & $0$      & $0$      & $-1$     & $0$         & $0$
    \\
    $\chi_{D^{(5)}}$    & $5$  & $5$   & $-1$     & $1$      & $1$      & $0$      & $-1$        & $-1$
    \\
    $\chi_{D^{(6)}}$    & $6$  & $-6$  & $\vdots$ & $0$      & $0$      & $1$      & $-\sqrt{2}$ & $\sqrt{2}$
    \\
    $\chi_{D^{(7)}}$    & $7$  & $7$   &          & $\vdots$ & $\vdots$ & $1$      & $-1$        & $-1$
    \\
    $\chi_{D^{(8)}}$    & $8$  & $-8$  &          &          &          & $0$      & $0$         & $0$
    \\
    $\chi_{D^{(9)}}$    & $9$  & $9$   &          &          &          & $\vdots$ & $1$         & $1$
    \\
    $\chi_{D^{(10)}}$ & $10$ & $-10$ &          &          &          &          & $\sqrt{2}$  & $-\sqrt{2}$
    \\
    $\chi_{D^{(11)}}$ & $11$ & $11$  &          &          &          &          & $\vdots  $  & $\vdots$
  \end{tabular}
  \caption{Characters for the reducible $\binoct$ representations showing the periodic columns up to the point where two rows repeat in order.}
  \label{tbl:reducible-character}
\end{table}

\section{Spin Wigner functions}

The graphical representations of spin operators used in this manuscript are analogous to the Wigner functions so often used to illustrate harmonic-oscillator operators.
As the set of spin coherent states forms a sphere in contrast to the plane of harmonic-oscillator coherent states, the functions representing the spin operators are functions on the sphere.
The map $W:A\mapsto W_A$ from operator to function satisfies the Stratonovitch-Weyl postulates:
\begin{itemize}
  \item Linearity: $W$ is linear and one-to-one
  \item Reality: $W_{A^\dagger}(\hat{\mathbf{n}})=W_A^*(\hat{\mathbf{n}})$
  \item Standardization: $4\pi\tr(A)=(2j+1)\int_{S^2}d\hat{\mathbf{n}}W_A(\hat{\mathbf{n}})$
  \item Traciality: $4\pi\tr(AB)=(2j+1)\int_{S^2}d\hat{\mathbf{n}}W_A(\hat{\mathbf{n}})W_B(\hat{\mathbf{n}})$
  \item Covariance: $W_{D(g)AD(g)^\dagger}(\hat{\mathbf{n}})=W_A(R^{-1}(g)\hat{\mathbf{n}})$, $g\in\SU(2)$
\end{itemize}
Here $R(g)$ is the $\SO(3)$ representation of the $\SU(2)$ group element $g$, which acts on the three-dimensional real vector space containing the unit vectors $\hat{\mathbf{n}}$ which make up the sphere $S^2$.
As shown in~\cite{stratonovich_distributions_1957,heiss_discrete_2000}, such a map $W$ is realized by taking the trace with a kernel $\Delta(\hat{\mathbf{n}})$ such that $W_A(\hat{\mathbf{n}})=\tr(\Delta(\hat{\mathbf{n}})A)$, where
\begin{align}
  \Delta(R(g)\hat{\mathbf{z}})
  &=
  \sum_{m=-j}^j\sum_{\ell=0}^{2j}\frac{2\ell+1}{2j+1}\iprodlr{
  \begin{matrix}j & \ell \\ m & 0\end{matrix}}{\begin{matrix}j \\ m\end{matrix}}
  D(g)\oprod{j,m}{j,m}D(g)^\dagger\,.
\end{align}
This is not the only choice of kernel satisfying the postulates, but it is a pleasing choice due to similarities to the parity operator discussed in~\cite{heiss_discrete_2000}.

\section{Support of codewords}

The requirement that $\exp(-i\tfrac{\pi}{2}J_z)$ yields $\pm\SL$ when restricted to the codespace means $\supp_{J_z}\ketL{a}\subseteq m_a+4\mathbf{Z}$ for $a\in\{0,1\}$, since $\SL$ only puts a phase on the computational-basis states.
The phases imparted, which depend on the state-dependent offset $m_a$, are $\SL\ketL{a}=\exp(-i\tfrac{\pi}{2}m_a)\ketL{a}$.
For these to yield the appropriate relative phases further imposes $m_0-m_1\equiv1\mod4$.
One obtains another constraint by recalling that $\exp(-i\pi J_x)$ performs a logical $X$ gate, exchanging $\ketL{0}$ and $\ketL{1}$.
Since this $\pi$ rotation about the $x$ axis inverts the $z$ axis, it must be that $\supp_{J_z}\ketL{0}=-\supp_{J_z}\ketL{1}$.
This leaves the two possibilities for the $\ketL{0}$ support presented in the main text:
\begin{align}
  \supp_{J_z}\ketL{0}
  &\subseteq
  \begin{cases}
    \frac{1}{2}+4\mathbf{Z}
  \\
  -\frac{3}{2}+4\mathbf{Z}\,,
  \end{cases}
\end{align}
where again the term $4\mathbf{Z}$ indicates the set of all integer multiples of 4.
The octahedral symmetry of the codes implies that analogous statements hold for the supports of the $\sigXL$ and $\sigYL$ eigenstates in the $J_x$ and $J_y$ bases, respectively.

\section{Quadratic gates}

Here I demonstrate that the Hamiltonian parameters given in the main text for the $\CZ$ and $\TL$ gates give the desired evolution on the codespace.
I begin with the simpler $\TL$ calculation, which illustrates most of the procedures necessary for the slightly longer $\CZ$ calculation.
In all calculations I express the logical states as
\begin{align}
  \ketL{a}
  &=
  \sum_mc_{a,m}\tightket{j,m}.
\end{align}

Apply a general Hamiltonian containing $J_z$ and $J_z^2$ terms to the computational basis states to figure out parameters yielding a $\TL$ gate.
Start with the case where $\supp_{J_z}\ketL{0}\subseteq\tfrac{1}{2}+4\mathbf{Z}$.
\begin{align}
  \exp(-i\phi J_z)\exp(-i\theta J_z^2)\ketL{a}
  &=
  \sum_kc_{a,4k\pm1/2}\exp\big({-}i(4k\pm\tfrac{1}{2})\phi-i(4k\pm\tfrac{1}{2})^2\theta\big)\ket{j,4k\pm\tfrac{1}{2}}
\end{align}
This exponential must not depend on $k$.
Write the coefficient of $-i$ in the exponent:
\begin{align}
  16\theta k^2+4(\phi\pm\theta)k+\tfrac{1}{4}(\theta\pm2\phi)\,.
\end{align}
Remove the quadratic term (modulo $2\pi$) by requiring $\theta=n\tfrac{\pi}{8}$:
\begin{align}
  (4\phi\pm n\tfrac{\pi}{2})k+n\tfrac{\pi}{32}\pm\tfrac{1}{2}\phi\,.
\end{align}
Remove the linear term (modulo $2\pi$) by setting $n=2$ ($\theta=\tfrac{\pi}{4}$) and $\phi=\tfrac{\pi}{4}$.
This leaves $\tfrac{\pi}{16}\pm\tfrac{\pi}{8}$ in the exponent.
The constant term is an overall phase on the code subspace which can be ignored, and the $\pm$ term gives precisely the relative phase difference between $\ketL{0}$ and $\ketL{1}$ needed to perform a $\TL$ gate.
An analogous calculation shows that $\theta=\tfrac{\pi}{4}$ and $\phi=\tfrac{5\pi}{4}$ implement a $\TL$ gate in the case where $\supp_{J_z}\ketL{0}\subseteq-\tfrac{3}{2}+\mathbf{Z}$.

Apply a general Hamiltonian containing $\tight{\Id\otimes J_z}$, $\tight{J_z\otimes\Id}$, and $\tight{J_z\otimes J_z}$ terms to the computational basis states to figure out parameters yielding a $\CZ$ gate.
Start with the case where $\supp_{J_z}\ketL{0}\subseteq\tfrac{1}{2}+4\mathbf{Z}$.
\begin{multline}
  \exp(-i\phi_1\tight{J_z\otimes\Id})
  \exp(-i\phi_2\tight{\Id\otimes J_z})
  \exp(-i\theta\tight{J_z\otimes J_z})\ketL{a}\ketL{b}
  =
  \\
  \sum_{k_1,k_2}c_{a,4k_1\pm_1 1/2}c_{b,4k_2\pm_2 1/2}
  \exp\big({-}i(\tight{4k_1\pm_1\tfrac{1}{2}})\phi_1-i(\tight{4k_2\pm_2\tfrac{1}{2}})\phi_2
  -i\theta(\tight{4k_1\pm_1\tfrac{1}{2}})(\tight{4k_2\pm_2\tfrac{1}{2}})\big)
  \tightket{j,4k_1\pm_1\tfrac{1}{2}}\tightket{j,4k_2\pm_2\tfrac{1}{2}}
\end{multline}
Again write the coefficient of $-i$ in the exponent:
\begin{align}
  16\theta k_1k_2+2(2\phi_1\pm_2\theta)k_1+2(2\phi_2\pm_1\theta)k_2\pm_1\tfrac{1}{2}\phi_1\pm_2\tfrac{1}{2}\phi_2
  \pm_1\!\pm_2\tfrac{1}{4}\theta\,.
\end{align}
Again require $\theta=n\tfrac{\pi}{8}$:
\begin{align}
  (4\phi_1\pm_2n\tfrac{\pi}{4})k_1+(4\phi_2\pm_1n\tfrac{\pi}{4})k_2\pm_1\tfrac{1}{2}\phi_1\pm_2\tfrac{1}{2}\phi_2
  \pm_1\!\pm_2n\tfrac{\pi}{32}\,.
\end{align}
To kill the linear terms, set $n=4q$ and $\phi=r\tfrac{\pi}{2}+q\tfrac{\pi}{4}$.
\begin{align}
  \pm_1\tfrac{1}{2}(r_1\tfrac{\pi}{2}+q\tfrac{\pi}{4})\pm_2\tfrac{1}{2}(r_2\tfrac{\pi}{2}+q\tfrac{\pi}{4})
  \pm_1\!\pm_2q\tfrac{\pi}{8}\,.
\end{align}
Take $r_1=r_2=r$.
For $\CZ$, the $+_1+_2$, $+_1-_2$, and $-_1+_2$ cases should all return the same phase.
The cross terms $+_1-_2$ and $-_1+_2$ yield the coefficient $-q\tfrac{\pi}{8}$.
Equating to the $+_1+_2$ term:
\begin{align}
  r\tfrac{\pi}{2}+q\tfrac{\pi}{4}+q\tfrac{\pi}{8}
  &\equiv
  -q\tfrac{\pi}{8}\mod2\pi\,,
\end{align}
implying $(r+q)\tfrac{\pi}{2}\equiv0\mod2\pi$, so $r+q$ must be a multiple of $4$.
To get the right phase difference in the $-_1-_2$ case it must be that
\begin{align}
  -r\tfrac{\pi}{2}-q\tfrac{\pi}{4}+q\tfrac{\pi}{8}
  &\equiv
  -q\tfrac{\pi}{8}+\pi\mod2\pi\,,
\end{align}
implying $-r\tfrac{\pi}{2}\equiv\pi\mod2\pi$, so $r$ must be twice an odd number.
Satisfy all criteria by choosing $r=q=2$, implying $\theta=\pi$ and $\phi=-\tfrac{\pi}{2}$.
An analogous calculation shows that the same parameters work in the case where $\supp_{J_z}\ketL{0}\subseteq-\tfrac{3}{2}+4\mathbf{Z}$.


\begin{table*}
$$
\begin{array}{c|rrrrrrr}
  \rm class & \rm1 & \rm2 & \rm3a & \rm3b & \rm4 & \rm6a & \rm6b\cr
  \rm size & 1 & 1 & 4 & 4 & 6 & 4 & 4\cr
\hline
  \chi_{\varrho_{1}} & 1 & 1 & 1 & 1 & 1 & 1 & 1\cr
  \chi_{\varrho_{2}} & 1 & 1 & e^{i4\pi/3} & e^{i2\pi/3} & 1 & e^{i4\pi/3} & e^{i2\pi/3}\cr
  \chi_{\varrho_{3}} & 1 & 1 & e^{i2\pi/3} & e^{i4\pi/3} & 1 & e^{i2\pi/3} & e^{i4\pi/3}\cr
  \chi_{\varrho_{4}} & 2 & -2 & -1 & -1 & 0 & 1 & 1\cr
  \chi_{\varrho_{5}} & 2 & -2 & e^{i5\pi/3} & e^{i\pi/3} & 0 & e^{i2\pi/3} & e^{i4\pi/3}\cr
  \chi_{\varrho_{6}} & 2 & -2 & e^{i\pi/3} & e^{i5\pi/3} & 0 & e^{i4\pi/3} & e^{i2\pi/3}\cr
  \chi_{\varrho_{7}} & 3 & 3 & 0 & 0 & -1 & 0 & 0\cr
\end{array}
\qquad
\begin{array}{c|rrrrrrrr}
  \rm class & \rm1 & \rm2 & \rm3 & \rm4a & \rm4b & \rm6 & \rm8a & \rm8b\cr
  \rm size & 1 & 1 & 8 & 6 & 12 & 8 & 6 & 6\cr
\hline
  \chi_{\varrho_{1}} & 1 & 1 & 1 & 1 & 1 & 1 & 1 & 1\cr
  \chi_{\varrho_{2}} & 1 & 1 & 1 & 1 & -1 & 1 & -1 & -1\cr
  \chi_{\varrho_{3}} & 2 & 2 & -1 & 2 & 0 & -1 & 0 & 0\cr
  \chi_{\varrho_{4}} & 2 & -2 & -1 & 0 & 0 & 1 & \sqrt{2} & -\sqrt{2}\cr
  \chi_{\varrho_{5}} & 2 & -2 & -1 & 0 & 0 & 1 & -\sqrt{2} & \sqrt{2}\cr
  \chi_{\varrho_{6}} & 3 & 3 & 0 & -1 & -1 & 0 & 1 & 1\cr
  \chi_{\varrho_{7}} & 3 & 3 & 0 & -1 & 1 & 0 & -1 & -1\cr
  \chi_{\varrho_{8}} & 4 & -4 & 1 & 0 & 0 & -1 & 0 & 0\cr
\end{array}
$$
\caption{Character tables for $2\mathrm{T}$ (left) and $\binoct$ (right)}
\label{tbl:2T-2O-char-table}
\end{table*}

\begin{table*}
$$
\begin{array}{c|rrrrrrrrr}
  \rm class & \rm1 & \rm2 & \rm3 & \rm4 & \rm5a & \rm5b & \rm6 & \rm10a & \rm10b\cr
  \rm size & 1 & 1 & 20 & 30 & 12 & 12 & 20 & 12 & 12\cr
\hline
  \chi_{\varrho_{1}} & 1 & 1 & 1 & 1 & 1 & 1 & 1 & 1 & 1\cr
  \chi_{\varrho_{2}} & 2 & -2 & -1 & 0 & \frac{-1+\sqrt{5}}{2} & \frac{-1-\sqrt{5}}{2} & 1 & \frac{1+\sqrt{5}}{2} & \frac{1-\sqrt{5}}{2}\cr
  \chi_{\varrho_{3}} & 2 & -2 & -1 & 0 & \frac{-1-\sqrt{5}}{2} & \frac{-1+\sqrt{5}}{2} & 1 & \frac{1-\sqrt{5}}{2} & \frac{1+\sqrt{5}}{2}\cr
  \chi_{\varrho_{4}} & 3 & 3 & 0 & -1 & \frac{1-\sqrt{5}}{2} & \frac{1+\sqrt{5}}{2} & 0 & \frac{1+\sqrt{5}}{2} & \frac{1-\sqrt{5}}{2}\cr
  \chi_{\varrho_{5}} & 3 & 3 & 0 & -1 & \frac{1+\sqrt{5}}{2} & \frac{1-\sqrt{5}}{2} & 0 & \frac{1-\sqrt{5}}{2} & \frac{1+\sqrt{5}}{2}\cr
  \chi_{\varrho_{6}} & 4 & 4 & 1 & 0 & -1 & -1 & 1 & -1 & -1\cr
  \chi_{\varrho_{7}} & 4 & -4 & 1 & 0 & -1 & -1 & -1 & 1 & 1\cr
  \chi_{\varrho_{8}} & 5 & 5 & -1 & 1 & 0 & 0 & -1 & 0 & 0\cr
  \chi_{\varrho_{9}} & 6 & -6 & 0 & 0 & 1 & 1 & 0 & -1 & -1\cr
\end{array}
$$
\caption{Character table for $2\mathrm{I}$}
\label{tbl:2I-char-table}
\end{table*}

\begin{table*}
  \centering
  \begin{tabular}{ c c c c }
    $\SU(2)$-irrep dim. & $\varrho_4$ mult. & $\varrho_5$ mult. & $\varrho_6$ mult. \\
    \hline
    $12q+0$             & $2 q$             & $2 q$             & $2 q$             \\
    $12q+2$             & $2 q + 1$         & $2 q$             & $2 q$             \\
    $12q+4$             & $2 q$             & $2 q + 1$         & $2 q + 1$         \\
    $12q+6$             & $2 q + 1$         & $2 q + 1$         & $2 q + 1$         \\
    $12q+8$             & $2 q + 2$         & $2 q + 1$         & $2 q + 1$         \\
    $12q+10$            & $2 q + 1$         & $2 q + 2$         & $2 q + 2$
  \end{tabular}
  \qquad
  \begin{tabular}{ c c c c c }
    $\SU(2)$-irrep dim. & $\varrho_1$ mult. & $\varrho_2$ mult. & $\varrho_3$ mult. & $\varrho_7$ mult. \\
    \hline
    $12q+1$             & $q + 1$           & $q$               & $q$               & $3 q$             \\
    $12q+3$             & $q$               & $q$               & $q$               & $3 q + 1$         \\
    $12q+5$             & $q$               & $q + 1$           & $q + 1$           & $3 q + 1$         \\
    $12q+7$             & $q + 1$           & $q$               & $q$               & $3 q + 2$         \\
    $12q+9$             & $q + 1$           & $q + 1$           & $q + 1$           & $3 q + 2$         \\
    $12q+11$            & $q$               & $q + 1$           & $q + 1$           & $3 q + 3$
  \end{tabular}
  \caption{Multiplicities of all $2\mathrm{T}$ irreps that appear in half-integer (left) and integer (right) irreps of $\SU(2)$.
  The dimension of the $\SU(2)$ irrep is presented in the form $12q+2p$ or $12q+2p+1$, where $q$ is any non-negative integer and $0\leq p\leq5$.}
  \label{tbl:2T-irreps}
\end{table*}

\begin{table*}
  \centering
  \begin{tabular}{ c c c c }
    $\SU(2)$-irrep dim. & $\varrho_4$ mult. & $\varrho_5$ mult. & $\varrho_8$ mult.
    \\
    \hline
    $24q$  & $2q$      & $2q$   & $4q$
    \\
    $24q+2$  & $2q+1$  & $2q$   & $4q$
    \\
    $24q+4$  & $2q$    & $2q$   & $4q+1$
    \\
    $24q+6$  & $2q$    & $2q+1$ & $4q+1$
    \\
    $24q+8$  & $2q+1$  & $2q+1$ & $4q+1$
    \\
    $24q+10$  & $2q+1$ & $2q$   & $4q+2$
    \\
    $24q+12$  & $2q+1$ & $2q+1$ & $4q+2$
    \\
    $24q+14$  & $2q+1$ & $2q+2$ & $4q+2$
    \\
    $24q+16$  & $2q+1$ & $2q+1$ & $4q+3$
    \\
    $24q+18$  & $2q+2$ & $2q+1$ & $4q+3$
    \\
    $24q+20$ & $2q+2$  & $2q+2$ & $4q+3$
    \\
    $24q+22$ & $2q+1$  & $2q+2$ & $4q+4$
  \end{tabular}
  \qquad
  \begin{tabular}{ c c c c c c }
    $\SU(2)$-irrep dim. & $\varrho_1$ mult. & $\varrho_2$ mult. & $\varrho_3$ mult. & $\varrho_6$ mult. & $\varrho_7$ mult.
    \\
    \hline
    $24q+1$  & $q+1$ & $q$   & $2q$   & $3q$   & $3q$
    \\
    $24q+3$  & $q$   & $q$   & $2q$   & $3q+1$ & $3q$
    \\
    $24q+5$  & $q$   & $q$   & $2q+1$ & $3q$   & $3q+1$
    \\
    $24q+7$  & $q$   & $q+1$ & $2q$   & $3q+1$ & $3q+1$
    \\
    $24q+9$  & $q+1$ & $q$   & $2q+1$ & $3q+1$ & $3q+1$
    \\
    $24q+11$ & $q$   & $q$   & $2q+1$ & $3q+2$ & $3q+1$
    \\
    $24q+13$ & $q+1$ & $q+1$ & $2q+1$ & $3q+1$ & $3q+2$
    \\
    $24q+15$ & $q$   & $q+1$ & $2q+1$ & $3q+2$ & $3q+2$
    \\
    $24q+17$ & $q+1$ & $q$   & $2q+2$ & $3q+2$ & $3q+2$
    \\
    $24q+19$ & $q+1$ & $q+1$ & $2q+1$ & $3q+3$ & $3q+2$
    \\
    $24q+21$ & $q+1$ & $q+1$ & $2q+2$ & $3q+2$ & $3q+3$
    \\
    $24q+23$ & $q$   & $q+1$ & $2q+2$ & $3q+3$ & $3q+3$
  \end{tabular}
  \caption{Multiplicities of all $\binoct$ irreps that appear in half-integer (left) and integer (right) irreps of $\SU(2)$.
  The dimension of the $\SU(2)$ irrep is presented in the form $24q+2p$ or $24q+2p+1$, where $q$ is any non-negative integer and $0\leq p\leq11$.}
  \label{tbl:2O-irreps}
\end{table*}

\begin{table*}
  \centering
  \begin{tabular}{ c c c c c }
    $\SU(2)$-irrep dim. & $\varrho_2$ mult. & $\varrho_3$ mult. & $\varrho_7$ mult. & $\varrho_9$ mult. \\
    \hline
    $60q+0$             & $2 q$             & $2 q$             & $4 q$             & $6 q$             \\
    $60q+2$             & $2 q + 1$         & $2 q$             & $4 q$             & $6 q$             \\
    $60q+4$             & $2 q$             & $2 q$             & $4 q + 1$         & $6 q$             \\
    $60q+6$             & $2 q$             & $2 q$             & $4 q$             & $6 q + 1$         \\
    $60q+8$             & $2 q$             & $2 q + 1$         & $4 q$             & $6 q + 1$         \\
    $60q+10$            & $2 q$             & $2 q$             & $4 q + 1$         & $6 q + 1$         \\
    $60q+12$            & $2 q + 1$         & $2 q$             & $4 q + 1$         & $6 q + 1$         \\
    $60q+14$            & $2 q + 1$         & $2 q + 1$         & $4 q + 1$         & $6 q + 1$         \\
    $60q+16$            & $2 q$             & $2 q$             & $4 q + 1$         & $6 q + 2$         \\
    $60q+18$            & $2 q$             & $2 q + 1$         & $4 q + 1$         & $6 q + 2$         \\
    $60q+20$            & $2 q + 1$         & $2 q + 1$         & $4 q + 1$         & $6 q + 2$         \\
    $60q+22$            & $2 q + 1$         & $2 q$             & $4 q + 2$         & $6 q + 2$         \\
    $60q+24$            & $2 q + 1$         & $2 q + 1$         & $4 q + 2$         & $6 q + 2$         \\
    $60q+26$            & $2 q + 1$         & $2 q + 1$         & $4 q + 1$         & $6 q + 3$         \\
    $60q+28$            & $2 q$             & $2 q + 1$         & $4 q + 2$         & $6 q + 3$         \\
    $60q+30$            & $2 q + 1$         & $2 q + 1$         & $4 q + 2$         & $6 q + 3$         \\
    $60q+32$            & $2 q + 2$         & $2 q + 1$         & $4 q + 2$         & $6 q + 3$         \\
    $60q+34$            & $2 q + 1$         & $2 q + 1$         & $4 q + 3$         & $6 q + 3$         \\
    $60q+36$            & $2 q + 1$         & $2 q + 1$         & $4 q + 2$         & $6 q + 4$         \\
    $60q+38$            & $2 q + 1$         & $2 q + 2$         & $4 q + 2$         & $6 q + 4$         \\
    $60q+40$            & $2 q + 1$         & $2 q + 1$         & $4 q + 3$         & $6 q + 4$         \\
    $60q+42$            & $2 q + 2$         & $2 q + 1$         & $4 q + 3$         & $6 q + 4$         \\
    $60q+44$            & $2 q + 2$         & $2 q + 2$         & $4 q + 3$         & $6 q + 4$         \\
    $60q+46$            & $2 q + 1$         & $2 q + 1$         & $4 q + 3$         & $6 q + 5$         \\
    $60q+48$            & $2 q + 1$         & $2 q + 2$         & $4 q + 3$         & $6 q + 5$         \\
    $60q+50$            & $2 q + 2$         & $2 q + 2$         & $4 q + 3$         & $6 q + 5$         \\
    $60q+52$            & $2 q + 2$         & $2 q + 1$         & $4 q + 4$         & $6 q + 5$         \\
    $60q+54$            & $2 q + 2$         & $2 q + 2$         & $4 q + 4$         & $6 q + 5$         \\
    $60q+56$            & $2 q + 2$         & $2 q + 2$         & $4 q + 3$         & $6 q + 6$         \\
    $60q+58$            & $2 q + 1$         & $2 q + 2$         & $4 q + 4$         & $6 q + 6$
  \end{tabular}
  \qquad
  \begin{tabular}{ c c c c c c }
    $\SU(2)$-irrep dim. & $\varrho_1$ mult. & $\varrho_4$ mult. & $\varrho_5$ mult. & $\varrho_6$ mult. & $\varrho_8$ mult. \\
    \hline
    $60q+1$             & $q + 1$           & $3 q$             & $3 q$             & $4 q$             & $5 q$             \\
    $60q+3$             & $q$               & $3 q + 1$         & $3 q$             & $4 q$             & $5 q$             \\
    $60q+5$             & $q$               & $3 q$             & $3 q$             & $4 q$             & $5 q + 1$         \\
    $60q+7$             & $q$               & $3 q$             & $3 q + 1$         & $4 q + 1$         & $5 q$             \\
    $60q+9$             & $q$               & $3 q$             & $3 q$             & $4 q + 1$         & $5 q + 1$         \\
    $60q+11$            & $q$               & $3 q + 1$         & $3 q + 1$         & $4 q$             & $5 q + 1$         \\
    $60q+13$            & $q + 1$           & $3 q + 1$         & $3 q$             & $4 q + 1$         & $5 q + 1$         \\
    $60q+15$            & $q$               & $3 q + 1$         & $3 q + 1$         & $4 q + 1$         & $5 q + 1$         \\
    $60q+17$            & $q$               & $3 q$             & $3 q + 1$         & $4 q + 1$         & $5 q + 2$         \\
    $60q+19$            & $q$               & $3 q + 1$         & $3 q + 1$         & $4 q + 2$         & $5 q + 1$         \\
    $60q+21$            & $q + 1$           & $3 q + 1$         & $3 q + 1$         & $4 q + 1$         & $5 q + 2$         \\
    $60q+23$            & $q$               & $3 q + 2$         & $3 q + 1$         & $4 q + 1$         & $5 q + 2$         \\
    $60q+25$            & $q + 1$           & $3 q + 1$         & $3 q + 1$         & $4 q + 2$         & $5 q + 2$         \\
    $60q+27$            & $q$               & $3 q + 1$         & $3 q + 2$         & $4 q + 2$         & $5 q + 2$         \\
    $60q+29$            & $q$               & $3 q + 1$         & $3 q + 1$         & $4 q + 2$         & $5 q + 3$         \\
    $60q+31$            & $q + 1$           & $3 q + 2$         & $3 q + 2$         & $4 q + 2$         & $5 q + 2$         \\
    $60q+33$            & $q + 1$           & $3 q + 2$         & $3 q + 1$         & $4 q + 2$         & $5 q + 3$         \\
    $60q+35$            & $q$               & $3 q + 2$         & $3 q + 2$         & $4 q + 2$         & $5 q + 3$         \\
    $60q+37$            & $q + 1$           & $3 q + 1$         & $3 q + 2$         & $4 q + 3$         & $5 q + 3$         \\
    $60q+39$            & $q$               & $3 q + 2$         & $3 q + 2$         & $4 q + 3$         & $5 q + 3$         \\
    $60q+41$            & $q + 1$           & $3 q + 2$         & $3 q + 2$         & $4 q + 2$         & $5 q + 4$         \\
    $60q+43$            & $q + 1$           & $3 q + 3$         & $3 q + 2$         & $4 q + 3$         & $5 q + 3$         \\
    $60q+45$            & $q + 1$           & $3 q + 2$         & $3 q + 2$         & $4 q + 3$         & $5 q + 4$         \\
    $60q+47$            & $q$               & $3 q + 2$         & $3 q + 3$         & $4 q + 3$         & $5 q + 4$         \\
    $60q+49$            & $q + 1$           & $3 q + 2$         & $3 q + 2$         & $4 q + 4$         & $5 q + 4$         \\
    $60q+51$            & $q + 1$           & $3 q + 3$         & $3 q + 3$         & $4 q + 3$         & $5 q + 4$         \\
    $60q+53$            & $q + 1$           & $3 q + 3$         & $3 q + 2$         & $4 q + 3$         & $5 q + 5$         \\
    $60q+55$            & $q + 1$           & $3 q + 3$         & $3 q + 3$         & $4 q + 4$         & $5 q + 4$         \\
    $60q+57$            & $q + 1$           & $3 q + 2$         & $3 q + 3$         & $4 q + 4$         & $5 q + 5$         \\
    $60q+59$            & $q$               & $3 q + 3$         & $3 q + 3$         & $4 q + 4$         & $5 q + 5$
  \end{tabular}
  \caption{Multiplicities of all $2\mathrm{I}$ irreps that appear in half-integer (left) and integer (right) irreps of $\SU(2)$.
  The dimension of the $\SU(2)$ irrep is presented in the form $60q+2p$ or $60q+2p+1$, where $q$ is any non-negative integer and $0\leq p\leq29$.}
  \label{tbl:2I-irreps}
\end{table*}

\end{document}